\documentclass[fleqn,usenatbib]{mnras}

\usepackage[T1]{fontenc}
\usepackage[british]{babel}
\usepackage{ae,aecompl}

\usepackage{graphicx}	% Including figure files
\usepackage{amsmath}	% Advanced maths commands
\usepackage{amssymb}	% Extra maths symbols
\usepackage{pgf}
\usepackage{siunitx}
\usepackage{ulem}
\usepackage{xspace}
\usepackage{tikz}
\usepackage{todo}
\usepackage{upgreek}
\usetikzlibrary{arrows.meta}
\usetikzlibrary{quotes,angles}
\usetikzlibrary{intersections}

\usepackage{newtxtext,newtxmath}
\usepackage{multirow}
\usepackage{orcidlink}

\usepackage{xpatch}
\makeatletter
\xpatchcmd\NAT@citex
 {%
  \@citea\NAT@hyper@{%
    \NAT@nmfmt{\NAT@nm}%
    \hyper@natlinkbreak{\NAT@aysep\NAT@spacechar}{\@citeb\@extra@b@citeb}%
    \NAT@date
  }%
 }
 {%
  \@citea
  \NAT@nmfmt{\NAT@nm}%
  \NAT@aysep\NAT@spacechar
  \NAT@hyper@{\NAT@date}%
 }
 {}{}
\xpatchcmd\NAT@citex
 {%
  \@citea\NAT@hyper@{%
    \NAT@nmfmt{\NAT@nm}%
    \hyper@natlinkbreak{\NAT@spacechar\NAT@@open\if*#1*\else#1\NAT@spacechar\fi}%
    {\@citeb\@extra@b@citeb}%
    \NAT@date
  }%
 }
 {
  \@citea
    \NAT@nmfmt{\NAT@nm}%
    \NAT@spacechar\NAT@@open\if*#1*\else#1\NAT@spacechar\fi
    \NAT@hyper@{\NAT@date}%
 }
 {}{}
\makeatother

\DeclareSIUnit\parsec{pc}
\DeclareUnicodeCharacter{2212}{-}

\newcommand{\dmu}[1]{\SI{#1}{\parsec\per\cubic\cm}}
\newcommand{\DM}{\text{DM}}
\definecolor{rvwvcq}{rgb}{0.08235294117647059,0.396078431372549,0.7529411764705882}
\newcommand{\psr}{PSR~J1713+0747\xspace}
\newcommand{\dlens}{d_\text{lens}}
\newcommand{\dpsr}{d_\text{psr}}
\newcommand{\veff}{v_\text{eff}}
\newcommand{\qty}[2]{\SI[mode=text]{#1}{#2}}
\newcommand{\phn}{\phantom{0}}

\title[Profile and DM events in \psr]{Profile changes associated with dispersion measure events in \psr}

\author[F. X. Lin et al.]{%
Fang Xi Lin$^{\orcidlink{0000-0002-6820-4275}}$\!,$^{1,2}$\thanks{Contact e-mail: \href{mailto:flin@cita.utoronto.ca}{flin@cita.utoronto.ca}}
Hsiu-Hsien Lin$^{\orcidlink{0000-0001-7453-4273}}$\!,$^{1}$
Jing Luo$^{\orcidlink{0000-0001-5373-5914}}$\!,$^{1}$
Robert Main$^{\orcidlink{0000-0002-7164-9507}}$\!,$^{4}$
James McKee$^{\orcidlink{0000-0002-2885-8485}}$\!,$^{1}$
\newauthor
Ue-Li Pen$^{\orcidlink{0000-0003-2155-9578}}$\!,$^{7,1,2,3,5,6}$
Dana Simard$^{\orcidlink{0000-0002-8873-8784}}$\!,$^{8}$
Marten H. van Kerkwijk$^{\orcidlink{0000-0002-5830-8505}}$$^{5}$
\\
$^{1}$Canadian Institute for Theoretical Astrophysics, University of Toronto, 60 St. George Street, Toronto, ON M5S 3H8, Canada\\
$^{2}$Department of Physics, University of Toronto, 60 St. George Street, Toronto, ON M5S 1A7, Canada\\
$^{3}$Dunlap Institute for Astronomy and Astrophysics, University of Toronto, 50 St. George Street, Toronto, ON M5S 3H4, Canada\\
$^{4}$Max-Planck-Institut f\"{u}r Radioastronomie, Auf dem H\"{u}gel 69, 53121, Bonn, Germany \\
$^{5}$Department of Astronomy and Astrophysics, University of Toronto, 50 St. George Street, Toronto, ON M5S 3H4, Canada\\
$^{6}$Canadian Institute for Advanced Research, 180 Dundas St West, Toronto, ON M5G 1Z8, Canada\\
$^{7}$Institute of Astronomy and Astrophysics, Academia Sinica,
11F of AS/NTU Astronomy-Mathematics Building, No.1, Sec. 4, Roosevelt Rd,\\ Taipei 10617, Taiwan, R.O.C.\\
$^{8}$ Cahill Center for Astronomy and Astrophysics, MC 249-17 California Institute of Technology, Pasadena, CA 91125, USA\\
}

\date{Accepted 2021 August 30. Received 2021 August 9; in original form 2021 June 17}

\pubyear{2021}

\begin{document}
\label{firstpage}
\pagerange{\pageref{firstpage}--\pageref{lastpage}}
\maketitle
\begin{abstract}
  Propagation effects in the interstellar medium and intrinsic profile changes can cause variability in the timing of pulsars, which limits the accuracy of fundamental science done via pulsar timing.
  One of the best timing pulsars, \psr, has gone through two `dip' events in its dispersion measure (DM) time series.
  If these events reflect real changes in electron column density, they should lead to multiple imaging.
  We show that the events are well-fit by an underdense corrugated sheet model, and 
  look for associated variability in the pulse profile using principal component analysis.
  We find that there are transient pulse profile variations, but they vary in concert with the dispersion measure, unlike what is expected from lensing due to a corrugated sheet.
  The change is consistent in shape across profiles from both the \textit{Green Bank} and \textit{Arecibo} radio observatories, and its amplitude appears to be achromatic across the \SI{820}{\mega\hertz}, \SI{1.4}{\giga\hertz}, and \SI{2.3}{\giga\hertz} bands, again unlike expected from interference between lensed images.
  This result is puzzling.
  We note that some of the predicted lensing effects would need higher time and frequency resolution data than used in this analysis.
  Future events appear likely, and storing baseband data or keeping multiple time-frequency resolutions will allow more in-depth study of propagation effects and hence improvements to pulsar timing accuracy.
\end{abstract}

\begin{keywords}
ISM: general; pulsars: individual: \psr
\end{keywords}

%%%%%%%%%%%%%%%%%%%%%%%%%%%%%%%%%%%%%%%%%%%%%%%%%%

%%%%%%%%%%%%%%%%% BODY OF PAPER %%%%%%%%%%%%%%%%%%

\section{Introduction}

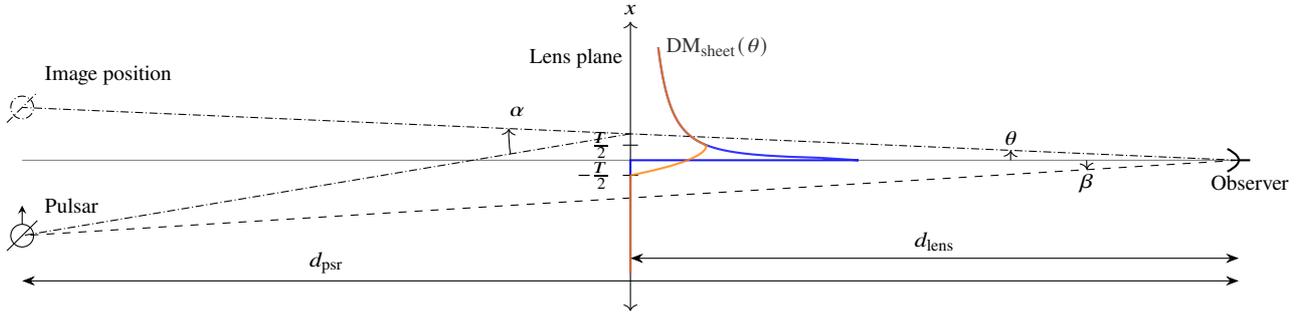
\begin{figure*}
\centering
\begin{tikzpicture}

% define coordinates
\coordinate (v1) at (0,-2);
\node (v2) at (0,2) {$x$};
\coordinate (obs) at (8,0);
\coordinate (v4) at (-8,0);
\coordinate (psr) at (-8,-1);
\coordinate (psrapp) at (-8,0.7);
\coordinate (T1) at (0,-0.2);
\coordinate (T2) at (0,0.2);

% draw axes
\draw [opacity=0.5] (obs) -- (v4);
\draw [<->, name path=lens plane] (v1) -- (v2) node[very near end, left] {Lens plane};

% draw optical lines and angles
\draw [dashed] (obs) -- (psr);
\draw [name path=apparent position, densely dash dot] (obs) -- (psrapp)
pic[draw, solid, <-, angle eccentricity=1.1, angle radius=3cm] {angle=psrapp--obs--v4}
pic[draw, solid, ->, angle eccentricity=1.1, angle radius=2cm] {angle=v4--obs--psr};
\draw [name intersections={of=lens plane and apparent position}, densely dash dot] (intersection-1) -- (psr) pic[draw, solid, <-, angle eccentricity=0.9, angle radius=1.6cm] {angle=psrapp--intersection-1--psr};
\node (alpha) at (-1.5,0.65) {$\alpha$};
\node (theta) at (5,0.3) {$\theta$};
\node (beta) at (6,-0.3) {$\beta$};

% draw the pulsar
\draw (psr) circle [radius=0.15cm];
\draw (psr) +(-0.2cm, -0.2cm) -- +(0.2cm, 0.2cm) node[above right] {Pulsar};
% draw velocity
\draw [-stealth](psr) +(0cm, 0.15cm)  -- +(0cm, 0.4cm);

% draw the pulsar apparent position
\draw [densely dash dot] (psrapp) circle [radius=0.15cm];
\draw [densely dash dot] (psrapp) +(-0.2cm, -0.2cm) -- +(0.2cm, 0.2cm) node[above right] {Image position};

% draw the observer
% \draw (0,0) parabola (1,1);
\draw [thick, rotate around={90:(obs)}] (obs) +(-0.15cm, 0.15cm) parabola bend (obs) +(0.15cm,0.15cm);
\draw [thick] (obs) -- +(0.15cm, 0) node[below=0.1cm] {Observer};

% define coordinates for distances
\coordinate (arrow origin 1) at (0,-1.3);
\coordinate (arrow origin 2) at (0,-1.6);
\coordinate (obs plane) at (8,-2);
\coordinate (lens plane) at (0,-2);
\coordinate (psr plane) at (-8,-2);

% draw distances
\draw[Stealth-Stealth] (arrow origin 1 -| obs plane) -- (arrow origin 1 -| lens plane) node[midway,above] {$d_\text{lens}$};
\draw[Stealth-Stealth] (arrow origin 2 -| obs plane) -- (arrow origin 2 -| psr plane) node[near end,above] {$d_\text{psr}$};

% draw DM model
\draw (T1) +(0.10cm,0) -- ++(-0.10cm,0) node[left=0.05cm] {$-\frac{T}{2}$};
\draw (T2) +(0.10cm,0) -- ++(-0.10cm,0) node[left=0.05cm] {$\frac{T}{2}$};

\draw [thick,color=blue,opacity=0.8] (0,-1.5) -- (0,0) -- (3,0) -- (2.5,0.032);
\draw [thick,domain=0.032:1.5,color=blue,opacity=0.8,samples=100] plot (-{-2*0.01*sqrt(1+1000/(2*\x))},\x);

\draw [thick,domain=0.22:1.5,color=orange,opacity=0.8,samples=100] plot (-{-2*0.01*sqrt(1+1000/(2*\x))},\x) node[right,color=black] {$\DM_\text{sheet} (\theta)$};
\draw [thick,domain=-0.2:0.18,color=orange,opacity=0.8,samples=100] plot (-{-(-6.25125*(\x-0.2)^2+2*0.01*sqrt(1+1000/(2*0.2)))},\x);
\draw [thick,domain=0.18:0.22,color=orange,opacity=0.8,samples=100] plot (-{-(-5.66*6.25125*(\x-0.18)^2+2*0.01*sqrt(1+1000/(2*0.2)))},\x);
\draw [thick,domain=-1.5:-0.2,color=orange,opacity=0.8] plot (0,\x);
\end{tikzpicture}
\caption{
  The lensing geometry we assume throughout \protect\citep[after fig. 2 b of][]{simardPredictingPulsarScintillation2018}.
  For simplicity, the folded sheet itself is not pictured.
  The horizontal axis is tangent to where the sheet is aligned with the direction to the observer.
  The functions $\DM_{\rm sheet} (\theta)$ sketch two possible column density profiles for the lens, with orange and blue for a thick and thin sheet, respectively, and $T$ the thickness of the unbent sheet (shown for the thick sheet). 
  For the thin sheet, $T$ is infinitesimally small, but its excess electron column density, and hence its DM contribution, is finite.
  The pulsar is located at an angle $\beta$ relative to the crest of the sheet, and its lensed image appears at an angle $\theta$ as its rays are bent by an angle $\alpha$ at the lens plane.
  The arrow on the pulsar shows its direction of motion.
}
\label{fig:sheetmodel}
\end{figure*}

The timing of millisecond pulsars (MSPs) has allowed for extremely precise measurements of masses and high-precision tests of general relativity \citep[e.g.][]{antoniadisMassivePulsarCompact2013,archibaldUniversalityFreeFall2018}.
The ensemble of well-timed pulsars also offers the hope of measuring very low-frequency gravitational waves \citep[e.g.][]{arzoumanianNANOGrav12Yr2020, verbiestPulsarTimingArray2021}.

Since pulsar timing is performed by correlating the observed profiles with a template profile that is assumed constant between observations, profile changes can bias times of arrival (TOAs).
While young pulsars have been observed to exhibit mode switching on month to decade time-scale \citep{lyneSwitchedMagnetosphericRegulation2010}, long-term profile changes are rarely observed in MSPs.
Recently, \cite{shannonDisturbanceMillisecondPulsar2016} reported a long-term profile change associated to a sudden magnetospheric shift in the MSP PSR~J1643--1224; \cite{mahajanModeChangingGiant2018} found mode changes on short time-scales in PSR~B1957+20; \citet{padmanabhRevisitingProfileInstability2021} has observed long-term profile instabilities in PSR~J1022+1001 that cannot be explained by miscalibration; and \citet{kerrParkesPulsarTiming2020} observed profile variations in PSR~J0437--4715, the closest and brightest MSP known, suggesting these profile changes may be in fact relatively common but unobserved.

Various propagation effects that arise in the interstellar medium (ISM) can also limit pulsar timing precision.
Propagation effects can cause scattering, scintillation broadening of pulses, and multiple imaging, which can all cause profile changes which impact TOAs, if not properly accounted for \citep[e.g.][]{stinebringEffectsInterstellarMedium2013,verbiestPulsarTimingArray2021}.
Conversely, understanding the propagation effects due to structures in the ISM should allow for the construction of better timing models and hence more sensitive timing arrays \citep{colesScatteringPulsarRadio2010,mainMeasuringInterstellarDelays2020a}.

Due to its narrow and high signal-to-noise (S/N) profiles, the 4.57-ms pulsar \psr is an excellent timer and is regularly observed by all four of the current pulsar timing arrays (PTAs): the \textit{North American Nanohertz Observatory for Gravitational Waves} \citep[\textit{NANOGrav}, ][]{alamNANOGrav12Yr2020}, the European Pulsar Timing Array \citep[][]{desvignesHighprecisionTiming422016}, the Parkes Pulsar Timing Array \citep[ ][]{kerrParkesPulsarTiming2020}, and the Indian Pulsar Timing Array \citep[][]{susobhananPintaUGMRTData2021}.
Recently, it has been used to put constraints on theories of gravity in the strong-field regime \citep{zhuTestingTheoriesGravitation2015,zhuTestsGravitationalSymmetries2019},
as well as to study single-pulse variability in detail \citep{liuVariabilityPolarimetryTiming2016}.
From these studies, it also appears that \psr exhibits uncorrelated epoch-to-epoch profile changes \citep{lentatiWidebandProfileDomain2017,brookNANOGrav11yearData2018}.
As precisions of PTAs continue to improve, our level of understanding of these profile changes will become a limiting factor in measurements that require precise timing, including in the detection of low-frequency gravitational waves.

An important component of pulsar timing models is a frequency-dependent delay term, 
characterized by the pulse arrival time relative to the infinite frequency arrival time
\begin{align}
    \label{eq:deltat}
    \Delta t \simeq \Delta t_\infty + \DM\frac{k_\text{DM}}{\nu^2},
\end{align}
where $\Delta t_\infty$ captures the frequency-independent terms, DM is the dispersion frequency (from hereafter DM), $\nu$ is the observing frequency, and
$k_\text{DM}=\text{c}r_\text{e}/2\uppi$, with $\text{c}$ the speed of light and $r_\text{e}$ the classical electron radius. In most timing packages, $k_\text{DM}$ is set equal to $10^4/2.41 \si{s.MHz^2.pc^{-1}.cm^3}$, which is thus the number that should be used to convert DM measurements to physical quantities. 
By measuring the relative delay at two different frequencies simultaneously, one can estimate DM, and therefore model out the dispersive delay (i.e. `dedisperse' the pulse).

The time delay arises due to the group delay of radio waves in the ionized ISM
\begin{align}
    \Delta t = \int_0^{\dpsr}\frac{\text{d}z}{v_\text{g}} = \int_0^{\dpsr}\frac{\text{d}z}{nc},
\end{align}
where $v_g=n \text{c}$ is the group velocity of light, and $\dpsr$ distance to the pulsar. For plasma, the refractive index is $n=\sqrt{1-2\text{c} k_\text{DM} n_\text{e}/\nu^2}$. For a typical electron density in the ionized ISM of $n_\text{e}\sim \qty{0.03}{cm^{-3}}$, the `plasma frequency' $2\text{c} k_\text{DM} n_\text{e} \sim \si{kHz}$ is much smaller than the typical observing frequency of $\sim \si{MHz}$. Thus, expanding $1/n$, the delay is
\begin{align}
    \label{eq:refractive_index_int}
    \Delta t \simeq \int_{0}^{\dpsr} \frac{\text{d}z}{\text{c}} \, \left( 1+\frac{\text{c}k_\text{DM} n_\text{e}}{\nu^2}\right).
\end{align}
Equating (\ref{eq:deltat}) and (\ref{eq:refractive_index_int}) gives that $\DM=\int_0^{\dpsr} n_\text{e}\, \text{d}z$. Because of this, DM tracks the free electron content along the line of sight to the pulsar, and changes in DM correspond to changes in the electron content and, as a result, changes in the index of refraction of the intervening plasma.

Because of the relative motion of Earth, the ISM, and the pulsar, DM changes over time as the path of the pulsar emission intersects different structures along the changing line of sight. Since DM is proportional to the refractive index difference from the vacuum, a large change in DM over a short time can lead to lensing events, such as the prominent echo observed in 1997 in the Crab pulsar \citep{backerPlasmaPrismModel2000}, and strong diffraction changes in PSR~J1603--7202 and PSR~J1017--7156 \citep{colesPulsarObservationsExtreme2015}.
Note that although the physical column density depends on two dimensions, it is only possible to sample DM along the motion of the pulsar.
DM variations can also be used to study the density structure of the ISM \citep[][]{phillipsTimeVariabilityPulsar1991}.

Observations of pulsar scintillation show that many pulsars have highly anisotropic and localized screens, which is inconsistent with a uniform and isotropic turbulent medium \citep[see e.g. ][]{putneyMultipleScintillationArcs2006,brisken100MasResolution2010}.
\citet{penPulsarScintillationsCorrugated2014} instead propose that pulsar scintillation is caused by thin sheets of a different electron density from the ambient ionized ISM that are closely aligned to the line of sight to the pulsar.
A small overdensity or underdensity of electrons within the sheet, when combined with small corrugations of the sheet and the orientation of sheet, can produce the large scattering angles observed in pulsars.
\citet{simardPredictingPulsarScintillation2018} developed a one-dimensional (1D) analytic description of the lensed images produced by a single corrugation on such a sheet, and the evolution of the images with time and observing frequency.

The DM time series of \psr shows two dips in $\sim\!\SI{13}{yr}$ of data \citep{jonesNANOGravNineyearData2017,lamSecondChromaticTiming2018}. Qualitatively, the dips look like the crossing of our line of sight through an underdense sheet, with a sharp dip followed by a slow recovery (see Fig.~\ref{fig:dms} below).
In this picture, the line of sight first moves through a region where the DM varies little, but then intersects a crest of a fold in the sheet leading to a sudden drop in DM, with a slow recovery as the effect of the sheet becomes less, back to roughly the same DM as at the start.

As previously mentioned, large DM changes should generically lead to multiple images.
For a thin underdense sheet, the geometry would be as shown in Fig.~\ref{fig:sheetmodel} \citep[see also figs~1 and~2 of][]{simardPredictingPulsarScintillation2018}, where emission directed in front of the observer gets deflected into the line of sight.
If the lensed ray has a delay larger than the pulse width, it will show up as an `echo', i.e. as a delayed copy of the pulse.
If, instead, the relative delay is less than the pulse width, it will interfere with the line of sight image and show up as a periodic modulation in the spectrum of the pulse \citep[e.g.][]{cordesMultipleImagingPulsars1986, backerPlasmaPrismModel2000, levkovPeriodicStructureFRB2020}.

In this paper, we quantitatively investigate the DM dips and look for profile evolution in the context of the above model.
The DM dips are well fitted by the model, which for the given distance to the lens give concrete predictions for time delays near the events.
We further find transient pulse shape variations (which we call `profile residuals' throughout the paper to avoid confusion with timing residuals) that are coincident with both DM events in \psr and that last roughly the same time as the events themselves, on the order of $\sim\!100$ d, but which we cannot reproduce in our model. This corroborates \citet{goncharovIdentifyingMitigatingNoise2021}, who, independently from this work, observed a profile change in \psr near the second DM event, but do not find such change near the first event.

During preparation of this manuscript, a profile change in \psr of similar nature but much larger amplitude than the one we observe was recorded by \citet{ATel14642Sustained}, and subsequently confirmed by \citet{ATel14652Confirmation}.
\citet{ATel14642Sustained} found that the event may be associated with a sharp change in DM, but \citet{ATel14652Confirmation} point out that a frequency-dependent profile change is often difficult to disentangle from a DM change.
It will be interesting to see whether the pulse profile and the DM will again recover to what they were before.

The paper is structured as follows: we first describe the data we use in section \ref{sec:data}, then detail the fit to the corrugated sheet model in section \ref{sec:sheetmodel}, and how we use principal component analysis (PCA) to find profile residuals in Section \ref{sec:pca}.
After a brief discussion of the impact on timing in section \ref{sec:timing}, we then  turn to potential causes of the profile change in section \ref{sec:discussion}, before concluding in section~\ref{sec:conclusion}.

\section{Data}
\label{sec:data}
\begin{figure*}
  \begin{center}
     \resizebox{\textwidth}{!}{\input{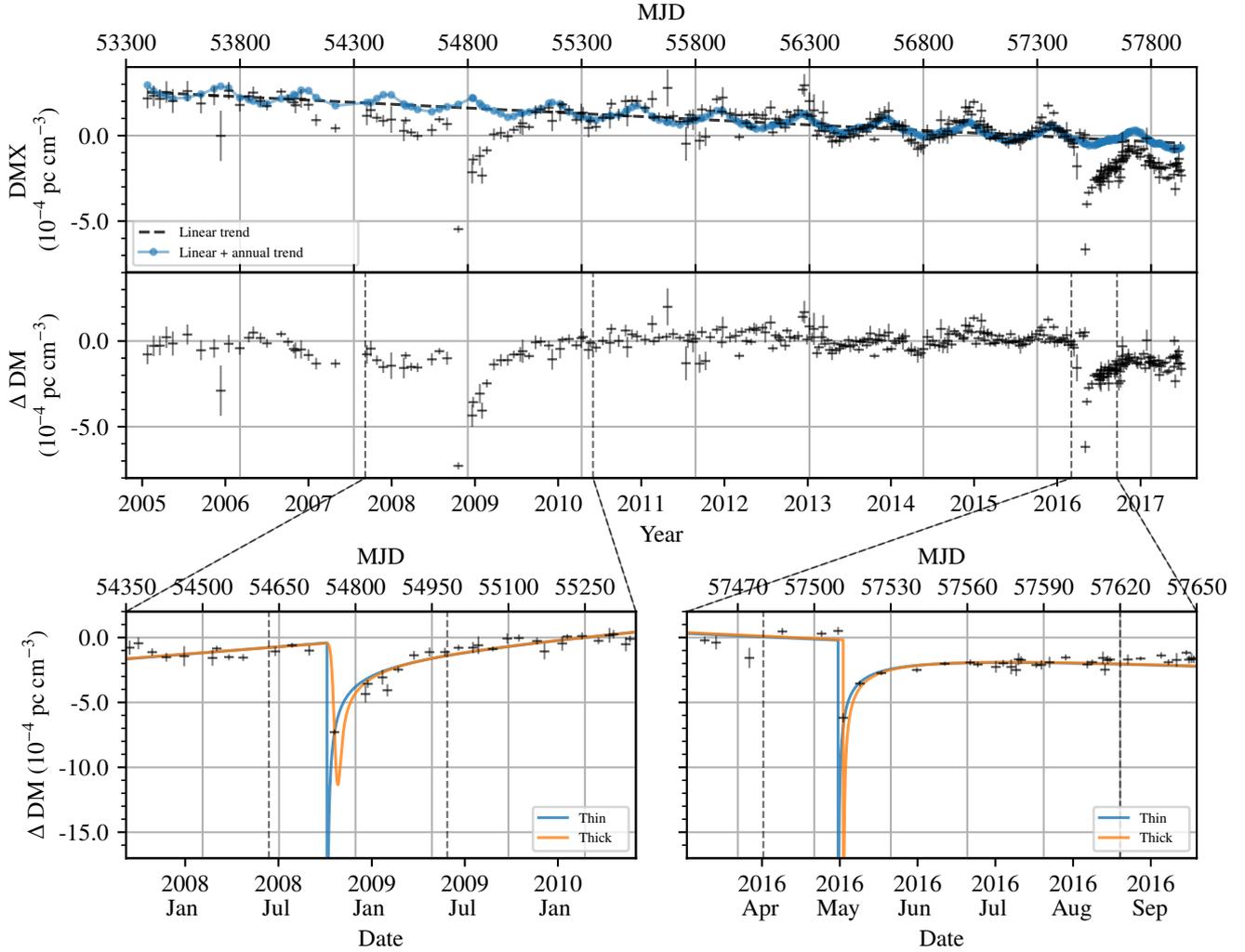}}
  \end{center}
  \caption{
    DM events in \psr.
    \textit{Top}: DMX from the \textit{NANOGrav} 12.5-yr data set.
    The dashed line is a fit of the linear trend and the points in blue are that fit plus an estimate of the annual modulation (see text).
    \textit{Middle}: Residuals $\Delta$DM.
    \textit{Bottom}: Enlargements of the $\Delta$DM profiles for both events.
    The drawn lines represent fitted $\Delta$DM profiles for one crest of an underdense sheet (see Section~\ref{sec:sheetmodel}).
    For both events, a `thin' and a `thick' sheet model is fit.
    The dashed vertical lines delimit the date range of the profiles displayed in Fig.~\ref{fig:profilesandresiduals}.
    }
    \label{fig:dms}
\end{figure*}
The \textit{NANOGrav} Collaboration monitors 77 MSPs with four telescopes, the 305-m \textit{William E. Gordon} (\textit{Arecibo}) \textit{Telescope of the Arecibo Observatory} (\textit{AO}), the 100-m \textit{Robert C. Byrd Green Bank Telescope} (\textit{GBT}) of the \textit{Green Bank Observatory}, the \textit{Karl G. Jansky Very Large Array}, and most recently, the \textit{Canadian Hydrogen Intensity Mapping Experiment} telescope \citep{ransomNANOGravProgramGravitational2019b}.
The data used in this paper come only from \textit{Arecibo} and \textit{GBT}.
The observations were of roughly monthly cadence until 2013 in \textit{GBT} and 2015 in \textit{Arecibo}, after which they were increased to a weekly cadence \citep{arzoumanianNANOGrav11yearData2018}.
Both telescopes received significant upgrades to their backends around 2010--12, between the first and second event.
The transition from ASP/GASP to PUPPI/GUPPI increased the available bandwidth from up to \SI{64}{\mega\hertz} to up to \SI{800}{\mega\hertz}.

We analysed the first \SI{11.5}{yr} of folded profiles for PSR J1713+0747 as a function of time, radio frequency, and polarization from \textit{NANOGrav}.
The profiles have 2048 phase bins, and frequency resolutions of either 1.5 MHz for PUPPI/GUPPI data, or 4 MHz for the older ASP/GASP data.
The pulse profile data are summarized in Table~\ref{tab:data}.

For our DM data, we take the excess DM (DMX) parameter from the \textit{NANOGrav} 12.5-yr data release\footnote{https://data.nanograv.org}.
This parameter models the time-dependent variation in DM as piece-wise constant during epochs of $\sim 6$ d or smaller ($\sim 15$ d in the older ASP/GASP data) relative to a fiducial DM, \dmu{15.9}, for \psr\citep[][]{arzoumanianNANOGrav11yearData2018}.
To measure DMX, \textit{NANOGrav} uses widely separated frequencies during each epoch, and excludes data with less than 10 per cent fractional bandwidth.

The DMX measurements for \psr have an overall decreasing trend as well as a significant annual variation due to a combination of Earth's motion and changes in the solar wind.
To account for these contributions as model independently as possible, we first subtract out a linear trend fit to DMX far from the dips (about 1.5 yr away on both sides).
We then fold the detrended DMX away from the dips to a 1-yr period, and apply a moving average filter with a window length of $\sim\SI{60}{\day}$ with periodic boundary to the folded DMX profile.
We subtract both the linear and annual trend, shown in blue in the top panel of Fig.~\ref{fig:dms} prior to fitting our model.

\begin{table}
  \centering
  \caption{Summary of the data used. The ASP/GASP data covered the first event (around MJD 54740), and the PUPPI/GUPPI data covered the second event (around MJD 57510).}
  \label{tab:data}
  \begin{tabular}{lccccr} % four columns, alignment for each
    \hline
              & Band  &         & Dates\\
    Telescope & (MHz) & Backend & (MJD)    & $N_{\rm obs}$\\
    \hline
    \hline
    AO\dotfill
    & 2316--2380 &        ASP   & 53394--56165 & \phn55\\
    & 1380--1444 &        ASP   & 53343--56165 & \phn78\\
    & 1700--2404 &        PUPPI & 55994--57609 &    105\\
    & 1147--1765 &        PUPPI & 55989--57609 &    125\\[0.8ex]
    GBT\dotfill
    &    1386--1434 &     GASP  & 53800--55639 & \phn62\\
    & \phn822--\phn866 &  GASP  & 53798--55641 & \phn62\\
    &    1151--1885 &     GUPPI & 55275--57607 &    215\\
    & \phn722--\phn919 &  GUPPI & 55278--57580 & \phn86\\
    \hline
  \end{tabular}
\end{table}

\section{Corrugated sheet model}
\label{sec:sheetmodel}
We briefly describe the lens model, and refer to \citet{simardPredictingPulsarScintillation2018} for details. Near the crest of a folded sheet, the equation of the sheet can be modelled as a parabola with an apex tangent to the line of sight
\begin{align}
    x=\frac{z^2}{2R},
\end{align}
where $x$ is the lens plane coordinate, $z$ is along the line of sight, and $R$ is the radius of curvature projected along the $x$ direction at the apex of the crest. Due to the fold, the column density integrated along $z$ near the apex $x=0$ is large (see Fig.~\ref{fig:sheetmodel}).
The column density profile of a thin sheet is
\begin{align}
    \DM (x) = 2\Delta n_\text{e} T\sqrt{1+\frac{R}{2x}}\qquad(x>0)
\end{align}
where $T$ denotes the thickness of the sheet, $\Delta n_\text{e}$ the constant excess electron density, and the pre-factor of 2 comes from the two sides of the parabola. For the thin sheet, $T$ is considered to be infinitesimally small, but the combination $\Delta n_\text{e} T$, and hence DM, finite.
A priori, we have no information about $R$, the radius of curvature of the sheet. Assuming parameters from~\citet{simardPredictingPulsarScintillation2018}, the radius of curvature $R$ is of order $\sim$ kpc, much larger than the spatial scale travelled by the pulsar in the data\footnote{The sheet itself does not extend that far, since it is typically wavy and folds back in the other direction. A large $R$ corresponds to a large radius of curvature of a single wave.} . Using angular units, $\theta=x/d_\text{lens}$ with $\dlens$ the distance from observer to the lens, and assuming that $\theta \ll R/2 d_\text{lens}$,
\begin{align}
    \label{eq:thin_lens}
    \DM_\text{thin} (\theta) \approx \sqrt{2} \frac{\Delta n_\text{e} T \sqrt{R}}{\dlens^{1/2}} \theta^{-\frac{1}{2}}\qquad(\theta>0),
\end{align}
with $\Delta n_\text{e} T \sqrt{R}$, a degenerate combination of $\Delta n_\text{e}$, $T$, and $R$ as the single lens parameter.
For a `thick' lens profile, the density in the sheet $\Delta n_\text{e} (d)$ depends on the depth $d$ through the unbent sheet. In that case, near the crest, the column density profile is approximately given by the convolution of $\Delta n_\text{e}$ and the profile of the lens 
\begin{equation}
    \label{eq:thick_eq}
    \DM_\text{thick} (\theta) = \int \text{d}\theta' \Delta n_\text{e}(\theta')\,2 \sqrt{1+\frac{R}{2\dlens(\theta'-\theta)}}.
\end{equation}
We will consider a Gaussian profile with full width at half-maximum of $T$ for $\Delta n_\text{e} (d)$ and total electron column density $\Delta n_{e,0} T$. When the images form far from the crest $|\theta|\gg|T/\dlens|$, the qualitative behavior of the images in the thin and thick lens models are similar.

We fit the underdense sheet model to the \textit{NANOGrav} DMX data by minimizing its $\chi^2$ in a section of data around the events.
In the first event, the crest is immersed in a background with significant variations in the DM, even after subtracting the mean annual variation.
We add a nominal linear function $\DM_0+\DM' \theta$ in the fit to account for this.
Additionally, we convert from the time-axis of the observed DMX to angles using the effective velocity $\theta=t\,\veff/\dlens$, where
\begin{align}
    \label{eq:effective_velocity}
    \textbf{v}_\text{eff} = (1-s)\,\textbf{v}_{\text{psr},\perp} + s\, \textbf{v}_{\text{obs},\perp} - \textbf{v}_{\text{lens},\perp},
\end{align}
with $s=1-\dlens/d_\text{psr}$ the fractional distance from the pulsar to the lens, $\textbf{v}_\text{psr}$ the pulsar velocity, $\textbf{v}_\text{obs}$ the observer velocity, and $\textbf{v}_\text{lens}$ the lens velocity, all transverse to the line of sight.
For a 1D model, there is an additional factor of $\cos\psi$, $\psi$ the angle between $\textbf{v}_\text{eff}$ and the lensing direction.

The pulsar velocity $v_\text{psr} = \qty{34.2}{km/s}$ using proper motion and distance from parallax is comparable, but larger than the Earth velocity $v_\text{obs} \lesssim \qty{31}{km/s}$.
Assuming a stationary lens, we thus get an upper bound for $\veff\cos\psi\leq v_\text{psr}$ from the triangle inequality on equation~(\ref{eq:effective_velocity}).
For the purpose of our analysis, we set $\veff\cos\psi = v_\text{psr}$ to give the `worst-case' bounds for observing lensing effect, giving a lower limit to delays and magnification of images, and an upper limit to the modulation bandwidth from multiple imaging.
Generally, DM events are more likely to be detected if the scattering/lensing direction is closely aligned with the pulsar's motion \citep[cf. scattering and proper motion alignment in PSR~B1508+55,][]{wucknitzImagingPulsarEchoes2019,marthiScintillationPSRB15082021a}.

\begin{table*}
  \centering
  \caption{Fitted parameters for the first and second event. For the thin lenses, the lens parameter is a product of the two columns $T$ and $\Delta n_\text{e}\sqrt{R}$. Two model selection criteria, $\chi_\text{reduced}^2$ and AIC$_\text{c}$ are also shown.
    }
  \label{tab:fits}
  \resizebox{\textwidth}{!}{
  \begin{tabular}{lcccc}
    \hline
    \hline
    &\multicolumn{2}{c}{\dotfill Event 1\dotfill}& \multicolumn{2}{c}{\dotfill Event 2\dotfill}\\
    Parameter & Thin lens & Thick lens & Thin lens & Thick lens\\
    \hline
    Date range (MJD)\dotfill
    & \multicolumn{2}{c}{54350--55350} &\multicolumn{2}{c}{57450--57600}\\
    Number of data points $N_{\rm data}$\dotfill
    & \multicolumn{2}{c}{37} & \multicolumn{2}{c}{23} \\[0.8ex]
    Crossing time $t_0$ (MJD)\dotfill
    & $54728\pm 5$ & $54761.2\pm 0.2$ & $57509.3\pm 0.5$ & $57511.35697\pm 0.0003$\\
    Width $T$ (\si{au})\dotfill
    & & $(1.33\pm0.04) \times 10^{-1}$ & & $(3.12\pm0.04) \times 10^{-4}$ \\
    Density term $\Delta n_\text{e}\sqrt{R}$ (\si{cm^{-3}.kpc^{1/2}})\dotfill
    & & $(-1.95\pm 0.05) \times 10^{-2}$ & & $-3.11\pm 0.09 $ \\
    Width-density product $T\Delta n_\text{e}\sqrt{R}$ (\si{au.cm^{-3}.kpc^{1/2}})\ldots
    & $(-3.8\pm 0.5) \times 10^{-3}$ & $(-2.61\pm0.09)\times10^{-3}$ & $(-1.3\pm 0.2) \times 10^{-3}$ & $(-9.7\pm 0.3)\times10^{-4}$\\
    Base DM $\DM_0$ (\si{10^{-5}.pc.cm^{-3}})\dotfill
    & $-4\pm 1$ & $-3.6\pm 0.5$ & $-2\pm 1$ &  $-0.02\pm 0.02$ \\
    Base DM gradient $\DM'$ (\si{10^{-7}.pc.cm^{-3}.d^{-1}})\dotfill
    & $3.2 \pm 0.3$ & $15.0\pm0.7$ & $-5 \pm 1$ & $-40\pm10$\\
    $\chi^2$\dotfill
    & 71.8 & 66.7 & 35.0 & 32.4\\
    $\chi_\text{reduced}^2$\dotfill
    & 2.18 & 2.08 & 1.84 & 1.80\\
    AIC$_\text{c}$\dotfill
    & 81.1 & 78.6 & 45.2 & 45.9\\
    \hline
  \end{tabular}
  }
\end{table*}

The full fitted function is then
\begin{align}
    \label{eq:thin_fitted}
    \DM_\text{thin} (\theta) = &\sqrt{2} \frac{\Delta n_\text{e} T \sqrt{R} }{v_\text{psr}^{1/2}} (t-t_0)^{-\frac{1}{2}} + \DM_0 + \DM'\cdot (t-t_0),
\end{align}
with $t_0$ a reference time corresponding to the epoch of the dip.
The fitted $\DM_\text{thick}$ is the first term in equation~(\ref{eq:thin_fitted}) convolved with the angle-dependent, gaussian thickness profile $\Delta n_\text{e}$ from equation~(\ref{eq:thick_eq}).
The presence of $T$ in the thickness profile partially breaks the degeneracy of the lens parameter $\Delta n_\text{e} T \sqrt{R}$.

In fitting a model $DM(\theta)$ to the observed DMs, and in the time-angular position conversion, we implicitly assumed that $DM(\theta)\simeq DM(\beta)$, since the observed DMs depend on the source position, and not the lens position.
In other words, we assume that the brightest image is never displaced by the lensing sufficiently for the measured DM to be different than the line of sight DM.
This can impact the quality of the fit, and a better fit may be achieved by iteratively fitting and solving for the lens equation. 
We choose to leave a more careful analysis for future studies.

The best-fitting model profiles are shown in the bottom panels of Fig.~\ref{fig:dms}, with the corresponding parameters, as well as the respective $\chi^2$ values, listed in Table~\ref{tab:fits}.
One sees that the model reproduces the data quite well, but the fit depends sensitively on how the single lowest DMX data point is interpreted.
For the thin sheet model, which starts with an abrupt change in DM, the lowest point  should happen in the tail, but for the thick sheet model, it can be on either side of the minimum.
In order to show the largest range in possibilities, we picked the case where the lowest point was at the start of the dip; on the other side, the thick sheet model looks very similar to the thin sheet one.
From Table~\ref{tab:fits}, one also sees that both the Akaike Information Criterion corrected for small sample size, AIC$_\text{c}$  \citep{liddleInformationCriteriaAstrophysical2007}, and the reduced $\chi^2$ prefer the thick model in the first event. AIC$_\text{c}$ marginally prefers the thin model in the second event, whereas the reduced $\chi^2$ marginally prefers the thick model.

\subsection{Lensing from the fold}
The lens equation for the thin lens (\ref{eq:thin_lens}) is
\begin{eqnarray}
\label{eq:lenseq}
    \theta &=& \beta - s\alpha(\theta)\\
  \alpha(\theta) &=& -\frac{\lambda^2 r_e}{2\uppi\dlens}\nabla_\theta \DM\nonumber\\
  &=& \frac{\lambda^2 r_e}{2\uppi}\frac{\Delta n_\text{e} T \sqrt{R}}{\sqrt{2}}\frac{1}{(\dlens \theta)^{3/2}}\quad(\theta>0),
\end{eqnarray}
where $\beta$ is the source position, $\alpha$ the bending angle (See Fig.~\ref{fig:sheetmodel}), $\lambda$ the observing wavelength, and $r_e$ the classical electron radius.
Lensing only depends on the gradient of the column density profile of the lens, i.e. the `geometry' of the lens.
Thus, the thick sheet model, as well as any other model with a sharp drop in column density followed by a slow recovery, gives qualitatively similar predictions for the evolution of the images\footnote{The thick model, being differentiable, obeys the odd-image theorem and produces a third image near the apex of its profile. However, the third image is very faint, and does not impact the results presented here.}.
We therefore focus on the lensing effects of a thin sheet as a prototype.

Images predicted by the model are solutions $\theta$ to the lens equation~(\ref{eq:lenseq}) for a given source position $\beta$.
The model predicts an evolution of the images as $\beta$ crosses the fold.
As shown in Fig.~\ref{fig:sheetmodel}, prior to the crossing, for pulsar angular position $\beta <0$, the sharp change of DM can refract the pulsar's signal, causing a lensed image in addition to the line of sight image.

Depending on the distance of the lens to the observer and the pulsar, different effects of the two images will be observed.
One generic expectation is that the second, lensed, image will manifest as an `echo', i.e., a copy of the pulse delayed in time, since the lensed image has a time delay $\tau$ with respect to the line of sight image, where
\begin{align}
    \label{eq:time_delay}
    \tau = \dpsr\frac{1-s}{s}\frac{\Delta\theta^2}{2c},
\end{align}
and $\Delta\theta$ the separation between the line of sight to the pulsar and lensed image.
The magnification of the image is given by,
\begin{equation}
    \label{eq:mag}
    M=\left(\frac{\partial\beta}{\partial\theta}\right)^{-1}
    =\left(1+s\frac{\lambda^2 r_e}{2\uppi}\frac{3}{2\sqrt{2}}\frac{\Delta n_\text{e} T\sqrt{R}}{(\dlens \theta)^{5/2}}\right)^{-1}.
\end{equation}
Another expectation is that the two images can interfere and will modulate in frequency due to the time delay and hence a phase delay inducing an amplitude modulation with bandwidth $\Delta \nu \sim 1/\tau$.

The angular separation is given by,
\begin{eqnarray}
    \label{eq:angular_separation}
    \Delta\theta &=& \theta-\beta \simeq -s\frac{\lambda^{2} r_e}{2\uppi}\frac{\Delta n_\text{e} T \sqrt{R}}{\sqrt{2}}\frac{1}{(\dlens\theta)^{3/2}} \nonumber\\
    &=&-\qty{1.0}{mas}\,\frac{s}{0.5}\left(\frac{\lambda}{\qty{0.21}{m}}\right)^2\left(\frac{\Delta n_\text{e} T \sqrt{R}}{-\qty{3.8e-3}{cm^{-3}.au.kpc^{1/2}}}\right)\nonumber\\
    &&\phantom{\qty{1.0}{mas}}\,\left(\frac{\dlens}{\qty{610}{pc}}\right)^{-3/2}\left(\frac{\theta}{\qty{0.3}{mas}}\right)^{-3/2},
\end{eqnarray}
where $\theta$ is the angular position of the lensed image, and where for the numerical estimate we scaled the lens distance to halfway to the pulsar.
The corresponding source positions is $\sim \qty{-0.7}{mas}$, corresponding to $\sim \qty{40}{d}$ before the dip using $\theta= \mu_\text{psr}t$, where $\mu_\text{psr}$ is the proper motion of the pulsar.
Inserting $\Delta\theta$ into equation~(\ref{eq:time_delay}) for the time delay, we find $\tau\sim \qty{1.5}{\us}$ for the lensed image, implying a modulation bandwidth of $\sim\qty{0.7}{MHz}$.
Inserting $\theta=\qty{0.3}{mas}$ into equation~(\ref{eq:mag}) for the magnification yields $M\simeq0.17$.
Hence, roughly $\qty{40}{d}$ prior to the dip, we expect another image, $\sim 0.17$ as bright, delayed by $\sim \qty{1.5}{\us}$, which evolves in time according to the lens equation (\ref{eq:lenseq}) and associated magnification (\ref{eq:mag}).
After the pulsar crosses over the dip, the model predicts that a single image remains.
At the crossing, the magnification of the remaining image should be analytically $2/5$, and afterwards it should recover asymptotically back to $1$, reaching $M \simeq0.8$ at $\sim\!\qty{50}{d}$.
Unfortunately, \textit{NANOGrav} only stores folded profiles with phase bins of $\SI{2.2}{\micro\second}$ and channel widths of 4 and \SI{1.5}{\mega\hertz} for the first and second event, respectively, insufficient to clearly resolve either the echo or the modulation in frequency.
An echo of a few percent in brightness at \SI{1}{\micro\second} can cause a slight broadening of the profile that evolves secularly in time over a time-scale of months prior to the dip, but we do not observe such an effect in the PCA as the following section shows.

However, the echo and amplitude modulations may be detectable in observing set-ups that store the baseband voltage data, which allows refolding of data to a desired time or frequency resolution.
The \textit{Large European Array for Pulsars} (\textit{LEAP}) observes \psr monthly since 2010 and stores the baseband data, thus may have captured the second event \citep{bassaLEAPLargeEuropean2016}.
An analysis of LEAP data could confirm or rule out our predictions.

\subsection{Robustness of lensing predictions}
\label{sec:lensing_predictions}
As discussed in the previous subsection, lensing models may have differences in their quantitative predictions of image positions and magnifications, but models sharing the same general geometry will share general qualitative predictions. In this section, we investigate the genericity of multiple imaging using the observed DM, independently from the sheet model.

For a given positive curvature in a locally underdense DM, a sufficiently low frequency, or a lens sufficiently close to Earth, will necessarily lead to multiple imaging due to focusing
\citep[cf.][]{cordesLensingFastRadio2017}. 
The condition for multiple imaging to occur is 
\begin{align}
    \label{eq:multiple_imaging}
    s\frac{\lambda^2 r_e}{2\uppi \dlens} \nabla_\theta^2 \DM \geq 1.
\end{align}
We estimate the curvatures near the two events from the DM data around the lowest points in DM, and bound the focal frequency and focal distance needed to get multiple imaging. 

For the first event, the curvature is $\qty{1.6e-7}{pc.cm^{-3}.d^{-2}}$. 
Converting to angular units with $\theta=\mu_\text{psr} t$, the curvature corresponds to $\qty{5.4e-4}{pc.cm^{-3}.mas^{-2}}$. 
Therefore, using equation~(\ref{eq:multiple_imaging}) and setting the lens to be half way from the pulsar, frequencies below the focal frequency of
\begin{align}
    \label{eq:focal_freq}
    f\lesssim \qty{870}{MHz} \left(\frac{s}{0.5}\right)^{1/2}\left(\frac{\dlens}{\qty{610}{pc}}\right)^{-1/2}\left(\frac{\nabla_\theta^2 \DM}{\qty{5.4e-4}{pc.cm^{-3}.mas^{-2}}}\right)^{1/2}
\end{align}
 must show multiple imaging. 
For the second event, we get a much larger curvature of  $\qty{4.4e-5}{pc.cm^{-3}.d^{-2}}$, or $\qty{0.15}{pc.cm^{-3}.mas^{-2}}$ with a corresponding focal frequency of \qty{14.4}{GHz}. 

We can also solve equation~(\ref{eq:multiple_imaging}) for the fractional distance, setting the frequency at $\qty{1.4}{GHz}$,
\begin{align}
    \label{eq:focal_dist}
    \frac{1}{s}-1 \lesssim 0.39 \left(\frac{f}{\qty{1.4}{GHz}}\right)^{-2}\left(\frac{\nabla_\theta^2 \DM}{\qty{5.4e-4}{pc.cm^{-3}.mas^{-2}}}\right),
\end{align}
so the focal fractional distance of the lens from the pulsar $s=1-\dlens/\dpsr\gtrsim 0.72$ in the first event. For the second event, a lens farther from the pulsar $s\gtrsim \qty{9.3e-3}{}$ will get multiple imaging due to focusing.

Note that the DM data in the first event are fairly sparse and likely underresolves the event in time, thus the curvature estimated from the DM data is a lower bound. Hence, equations~(\ref{eq:focal_freq}) and (\ref{eq:focal_dist}) are lower bound on the focal frequency and upper bound on focal fractional distance, respectively. We conclude that multiple imaging may be observable during the first, and is very likely observable during the second event.

We can also estimate the magnification once the pulsar crosses behind the dip from the curvature of DM.
For the first event, assuming the image to be near the first DM point after the minimum, the curvature is roughly \qty{-4.7e-8}{pc.cm^{-3}.d^{-2}}. Converting again to angular units, this corresponds to \qty{-1.6e-4}{pc.cm^{-3}.mas^{-2}}.
\begin{align}
    \label{eq:magnification_estimate}
    1-M^{-1} &=
    s\frac{\lambda^2 r_e}{2\uppi \dlens}\nabla^2_\theta \DM \nonumber\\
    &= -0.11\frac{s}{0.5}\left(\frac{\lambda}{\qty{0.21}{m}}\right)^2\left(\frac{\dlens}{\qty{610}{pc}}\right)^{-1}\nonumber\\
    &\phantom{=-0.11}\frac{\nabla^2_\theta \DM}{\qty{-1.6e-4}{pc.cm^{-3}.mas^{-2}}},
\end{align}
so that $M \sim 0.9$ at MJD $\sim 54825$.
For the second event, we find $M \sim 0.2$ at MJD $\sim 57525$.
One sees that these numbers are similar to what we inferred from the more detailed model; in order of magnitude they simply follow from the depression of the DM and the time-scale.
Similarly, given the negative curvature during the DM `recovery' after the dips, the underdense lens is guaranteed to be diverging, and hence the magnification has to be lower than the unity.

Note that so far we only considered the trailing edge of the dip.
There are even larger gradients in the leading edge, which will bend light towards post-crossing, leading to multiple images for the period after the DM dip. 
As discussed above, for the first event, gradient and curvature estimates are fairly uncertain due to the lack of data near the leading edge.
For the second, the sharp gradient will lead to a large bending angle of \qty{13}{mas} at \qty{1.4}{GHz} and $\dlens=\dpsr/2$, corresponding to a delay of \qty{0.25}{ms} using (\ref{eq:time_delay}), and the image will be dim with magnification $M \lesssim 0.01$.
Since it originates from a spatially compact region, the image will also be fairly achromatic, and would present itself as an outgoing `echo' in time. 
Comparing these with the predictions from the thick lens model, the image from the leading edge has $M\sim 0.03$ immediately post-dip and quickly decays to $M\sim0.0005$ after only 5 d.

\section{Profile variation analysis via PCA}
\label{sec:pca}
An echo of a few percent at a delay smaller than the time resolution of the data can be difficult to characterize.
Simply template matching two shifted and scaled copies of a template profile against the data is highly degenerate.
However, one expects that an echo will generically change the shape of the observed profile, and that the change should be near the DM events.
Hence, we look for associated changes.

For \psr, the analysis of profile changes is made more difficult by the fact that the integrated profiles are known to change slightly from observation to observation.
However, since we expect the perturbation to the profile to be small and localized in time compared to the overall data, we can use PCA to analyze the variability of the profile over the full time period, and then analyze the profile residuals over the times of interest.
PCA has been used as a way of mitigating timing effects due to profile variations in pulsar timing \citep{oslowskiHighSignaltonoiseRatio2011,linImprovedPulsarTiming2018} and to optimize the DM of giant pulses from Crab pulsar \citep{2021arXiv210513316T}.
For \psr, PCA was used to study single-pulse variations by \citet{shannonPulseIntensityModulation2012}, who found that it was sensitive even to subtle profile changes.

\begin{figure}
  \begin{center}
    \input{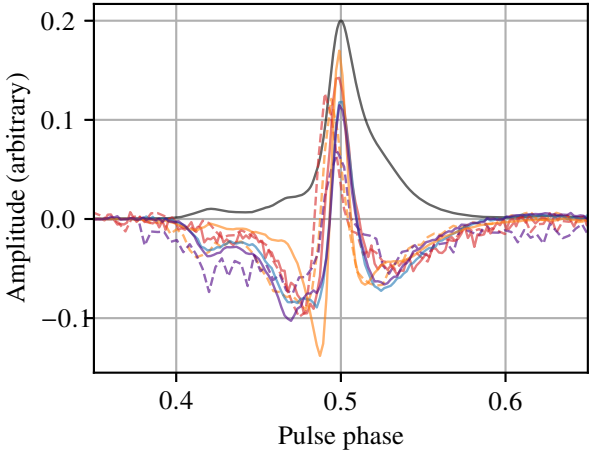}
  \end{center}
  \caption{
    The `broadening' mode of the pulse profile of \psr, found as the second most significant component in all pulse profile data sets except for the \textit{Green Bank} L-band GASP data.
    The red, blue, orange, and purple are the broadening mode profiles from \textit{Green Bank} at 820 MHz and 1.5 GHz, and \textit{Arecibo} at 1.4 GHz, and 2.3 GHz, respectively, with dashed and solid lines indicating ASP/GASP and PUPPI/GUPPI data, respectively.
    The average 1.4-GHz pulse profile is also shown in black.
    All profiles are rebinned for visual clarity.
    Note that the \textit{Arecibo} 1.4-GHz broadening mode appears to be sharper than that of the other data sets; this is due to the high S/N of the profiles.
    See Section~\ref{sec:pca} for discussion.
  }
  \label{fig:broadeningmode}
\end{figure}

\begin{figure*}
  \begin{center}
     \resizebox{\textwidth}{!}{\input{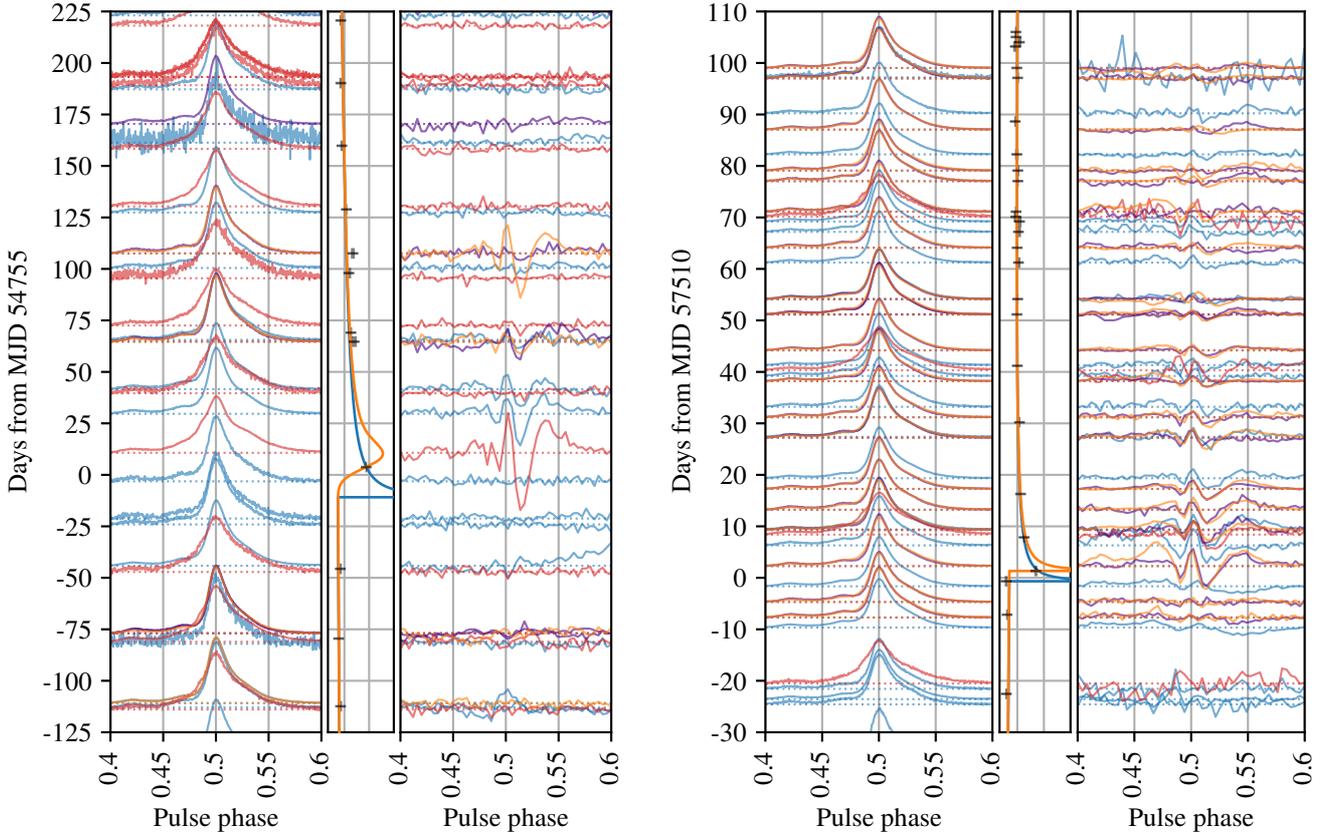}}
  \end{center}
  \caption{
    PSR J1713+0747 pulse profiles and residuals during the two DM events.
    For each event, the left-hand panel shows the profiles, the middle the DMX data and their fits from Fig.~\ref{fig:dms}, and the right-hand panel profile residuals.
    All profiles are rebinned by factor 8 in phase to reduce visual clutter.
    For the first event (left figure), to better show the shape of the profile residuals, they have been such that the off-pulse region of each residual profile has the same variance.
    For the second event (right figure), the profiles are of more uniform quality and hence we instead simply scaled all residuals by a factor 30 relative to the corresponding profiles, so that one can more easily see how the profile residuals decay in time.
    Colours are as in Fig~\ref{fig:broadeningmode}, with red, blue, orange, and purple used for \textit{Green Bank} at 80 0MHz and 1.5 GHz, and \textit{Arecibo} at 1.4 GHz and 2.3 GHz, respectively.
    In both events and at all frequencies and telescopes, we see a `W'-shaped feature appear after the DM `dip', which then decays away in time, indicating a similar change in pulse profile happened in both events.
    Note the difference in the plotted time-scale for the two events.
  }
  \label{fig:profilesandresiduals}
\end{figure*}

To determine the principal components of the pulse profile variations, we follow \cite{linImprovedPulsarTiming2018} and use PCA via singular value decomposition (SVD),
applying it separately on each of the eight sets of observations in Table~\ref{tab:data}.
We proceed as in section 3.1 of~\citet{linImprovedPulsarTiming2018}.
We first align the profiles from each observation to the template profiles provided by \textit{NANOGrav} using the typical fast Fourier transform timing technique outlined by \citet{taylorPulsarTimingRelativistic1992}.
In order to eliminate bias and spurious detection from the off-pulse region in the SVD analysis, we zero-out pulse phase 0--0.25 and 0.75--1.0.
For each set of observations, we produce a data matrix $D_{Ei}$ of frequency-averaged pulse profiles, where $E$ is the observation epoch index, and $i$ the pulse phase index.
Then, performing the SVD gives,
\begin{equation}
  D_{Ei} = (\textbf{U}\Sigma \textbf{V}^T)_{Ei} = \sum_{j} \textbf{U}_{Ej}\Sigma_{jj}\textbf{V}_{ij},
\end{equation}
where for each $j$, we can interpret the right-singular vectors $\textbf{V}_{ij}$ as the $j^\text{th}$ `mode' of the pulse profiles, the left-singular vectors $\textbf{U}_{Ej}$ as the relative amplitudes at epoch $E$ of mode $j$, and $\Sigma_{jj}$ the $j^\text{th}$ largest singular value.
In other words, the zeroth $V$-mode $\textbf{V}_{i0}$ corresponds to the mean pulse profile across the data set, the first mode $V$-mode $\textbf{V}_{i1}$ to the `largest change' from the mean pulse profile, and so forth.

With the exception of the \textit{Green Bank} L-band data of the first event, every set of observations also seems to show broadening or sharpening relative to the mean profile.
This `broadening mode' dominates the variation from the mean profile, and shows up typically\footnote{Particularly noisy profiles in the data set can sometimes cause $V$-modes with larger singular values. We ignore those modes.} in the second-strongest mode $\textbf{V}_{i1}$, with a peak at a phase of 0.5 and dip on either side of the peak.  Fig.~\ref{fig:broadeningmode} shows this mode for each of the data sets.
Epoch-to-epoch variation was previously seen in an analysis of profile stochasticity in PSR J1713+0747 using the Parkes telescope data at 732 MHz, 1369 MHz, and 3100 MHz, in ways consistent with our broadening mode \citep[cf. Fig.~10 in ][]{lentatiWidebandProfileDomain2017}.
In addition, the variation was independently observed in a subset of the \textit{NANOGrav} data set in \citet{brookNANOGrav11yearData2018}.

Since the variation is seen in multiple telescopes, it cannot be due to an instrumental factor.
The time-scale of the broadening change appear to be longer than a typical observation $\sim\qty{1}{h}$, but shorter than the typical interval between observations $\sim\qty{1}{week}$ \citep{brookNANOGrav11yearData2018}.
We also ruled out the broadening mode being due to scintillation by independently performing the SVD for different sub-bands of the data.
The broadening mode persists and appears to be consistent across sub-bands.
It is unclear whether the broadening is intrinsic to the pulsar or due to propagation effects.
We relegate a more in-depth study of the broadening effect to future studies.

Here, since the broadening mode is uncorrelated between epochs and thus not of interest for looking for changes associated with the DM dip, we will use only what is left after removal of the mean profile and the broadening mode (i.e. the zeroth and first $V$-modes).
In other words, we look at residuals defined by,
\begin{equation}
    R_{Ei} = D_{Ei} - \sum_{j=0,1} \textbf{U}_{Ej}\Sigma_{jj}\textbf{V}_{ij}.
\end{equation}
The profiles $D_{Ei}$ and profile residuals $R_{Ei}$ near the two DM dip events are shown in Fig.~\ref{fig:profilesandresiduals}, together with the $\Delta\DM$ data and fitted sheet models.

Before discussing the result, we note that by removing two components, the residuals become more difficult to interpret than would be the case if one just subtracted a static profile.
However, as can be seen in the right-hand panels of Fig.~\ref{fig:profilesandresiduals_meansubtraction} in the Appendix, the residuals relative to the mean profile are dominated by the previously discussed epoch-to-epoch broadening variations, and while profile variability associated with the DM event can still be discerned in the first event, it is swamped by the variations associated with the broadening mode in the second event.

We note also that for high enough S/N profiles, such as those from the wideband \textit{Arecibo} PUPPI-era data, a single \textit{V}-mode may not enough to describe the broadening mode.
Indeed, we find that multiple modes beyond the first two are not noise like.
This is not unexpected: SVD produces a linear decomposition of the data matrix $D_{Ei}$, but the physical processes causing changes in profiles are rarely linear in epoch and pulse phase.
To facilitate comparison between data sets, we have removed only up to two \textit{V}-modes in each of our sets of observations.
For completeness, we note though that residuals associated with the DM dip still appear when we remove up to the first $\sim 20$ modes in the \textit{Arecibo} L-band data (after which the residuals becomes difficult to distinguish from noise).

From Fig.~\ref{fig:profilesandresiduals}, one sees that the profile residuals $R_{Ei}$ are largest closely after either DM event, with a `W'-shaped feature, which decays away on a time-scale roughly similar to that of the DM recovery.
The feature appears in, and is consistent across, both \textit{Arecibo} and \textit{GBT} at all four observational frequencies.
It is especially clear for the second event, mostly because the observations at that time used a much higher bandwidth and thus led to much less noisy profiles.
The `W'-shaped feature is not linear in the data, and hence does not have a single associated mode.
Near pulse phase 0.5, the amplitude of the residual is $\sim\!1$ per cent of the peak profile amplitude.
Closer to pulse phase $0.54$, the residual has a `shoulder' with amplitude of $\sim\!4$ per cent of the profile amplitude at the same phase.

\section{Impact on timing}
\label{sec:timing}
Although we observe a clear profile change coincident with the DM events, it is difficult to independently measure the impact of the profile change on timing with our analysis, in part because we first align the profiles prior to taking the SVD, making any observed profile change symmetric.
Given the small amplitude of the profile change, however, the observed TOA residuals of $\lesssim\SI{0.1}{\micro\second}$ at the events are unlikely to be explained by the profile variation, but does place an upper bound on their impact.
Moreover, since the profile change appears to be roughly achromatic, it will not cause the chromatic TOA variations that are now interpreted as an excess in DM.

In the 1D corrugated sheet model, however, the line of sight image abruptly disappears after the dip and the leftover lensed image will be delayed because it follows a bent path.
The corresponding refractive geometric delay will scale as $\tau_\text{geo}\propto \nu^{-\xi}$, where the exact frequency scaling index $\xi$ depends on the shape of the column density profile.
Generically, one expects $\xi$ to be different from 2 and 4, the scaling indices for the regular dispersive delay and for a purely linear gradient lens (i.e. a DM `prism').
For a compact lens such as the thin sheet, the scaling can be from fairly shallow to very shallow.
Indeed, right behind the crossing, for $\beta=0$, one finds $\xi=8/5$ from the lens equation (\ref{eq:lenseq}), which may be difficult to disentangle from the $\tau\propto\nu^{-2}$ delay expected from dispersion. 
Experimentally, \citet{brisken100MasResolution2010} has identified very shallow ($\xi\lesssim 0.1$) frequency scaling of scattered images. 
In the timing analysis of the DM events by \citet{lamSecondChromaticTiming2018}, in which $\nu^{-2}$ and $\nu^{-4}$ delays were fit simultaneously, the $\nu^{-4}$ term appears to make some contribution near the dip, but it clearly does not dominate.

\section{Discussion}
\label{sec:discussion}
The systematic profile changes after the DM drops are a significant effect, even if they are small, of only a few percent depending on pulse phase.
The small magnitude of the effect may be the reason they were not seen in the previous analysis by \citet{lamSecondChromaticTiming2018}.
Near the second event, \citet{goncharovIdentifyingMitigatingNoise2021} find, in addition to a profile change similar to what we observe, that the ToAs near the second event follow a chromatic index $\xi=1.15$, and suggests that the event be of magnetospheric origin.
If the profile change arises from astrophysical causes, \psr would be the second MSP to show long-term temporal pulse profile changes, after PSR~J1643--1224 \citep{shannonDisturbanceMillisecondPulsar2016}. 

There are a number of possible reasons for profile variability, such as scatter broadening, echoes, scintillation, pulse jitter, unmodelled rotation measure causing leakage, inaccurate DM measurements, improper polarization calibration, or other instrumental and radio-frequency interference (RFI) effects \citep[see, e.g.][]{brookNANOGrav11yearData2018}.
These can be roughly grouped into artificial, intrinsic, and propagation effects.
Below, we discuss each of these three possibilities.

\subsection{Artificial effects}
\label{sec:artificialeffects}
The most obvious possibility is that since the DM suddenly jumps, the profiles were produced by dedispersing to an incorrect DM.
However, the amplitude of the profile residuals are of the same order at both \qty{820}{\MHz} and \qty{2.3}{\GHz}.
Dedispersion to a wrong DM before frequency averaging will lead to pulse broadening that is more pronounced at lower frequencies.
At most, a $\Delta\DM$ of \dmu{1e-3} can cause a delay of \SI{3}{\micro\second} between the top and bottom band at the lowest band, \qty{820}{\MHz} where dispersion would have the largest effect.
This is only slightly larger than one pulse phase bin of \SI{2.2}{\micro\second}.
This would produce a slightly broadened average profile, but not enough to account for the observed effect.

Other artificial effects can include instrumental, calibration, and RFI-related effects.
However, since the profile residuals have the same characteristics at two different telescopes and four different observating bands, and across major upgrades to both telescopes, we conclude that these effects cannot be the cause of the profile change.

\subsection{Intrinsic effects}
A well-known cause of intrinsic pulse variability is pulse jitter, typically expected to be uncorrelated between pulses.
Although it can cause pulse variation in \qty{30}{min}-long integrations, jitter does not cause systematic changes.
Pulse jitter in \psr has been extensively analysed \citep{shannonPulseIntensityModulation2012,liuVariabilityPolarimetryTiming2016,lentatiWidebandProfileDomain2017}.
Moreover, the appearance of the `W'-shaped profile residual coincides exactly with the drops in DM twice in 11.5 yr, thus the profile residual being caused by pulse jitter is unlikely.

Glitches can also cause changes in TOA, which might be mistaken for DM changes.
They are typically not associated with pulse profile changes \citep[see e.g.][]{espinozaStudy315Glitches2011}, and have only been observed in two MSPs: PSR~B1821--24A and PSR~J0613--0200 \citep{cognardMicroglitchMillisecondPulsar2004,mckeeGlitchMillisecondPulsar2016}.
Furthermore, the TOAs closely follow the $\nu^{-2}$ dispersion relation at both of the events, with some evidence for a $\nu^{-4}$ contributions \citep{lamSecondChromaticTiming2018}, while for a glitch or other magnetospheric effect one would generically expect an achromatic change.
Thus, we conclude that this is unlikely to explain the observed DM dips and profile changes.

The profile change could also be associated with magnetospheric changes.
In PSR J1643--1224, \citet{shannonDisturbanceMillisecondPulsar2016} reported a change sudden in TOAs potentially caused by a sudden magnetospheric shift.
In that case, the change in profile decays over a period of several months, similar to the change reported here, though of a much larger amplitude.
The TOAs also showed a sharp dip followed by slow recovery, somewhat similar to what might be inferred from the DM changes in \psr.
But for PSR J1643--1224 the TOA residuals were largest at higher frequencies, and they were of a magnitude commensurate with what one would expect from the profile changes, while here they are largest at lower frequency, following the $\nu^{-2}$ scaling expected from dispersion, and larger than would be expected taking the profile changes at face value.

\subsection{Propagation effects}
\begin{figure}
    \begin{center}
        \input{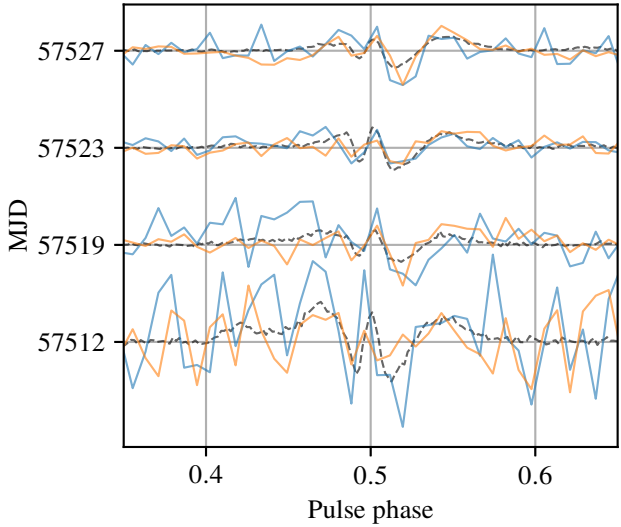}
    \end{center}
    \caption{
   Profile residuals after the second event at L-band at \textit{Arecibo}.
   One sees that the residuals obtained from single \SI{1.5}{\mega\hertz}-wide channels [at \SI{1200}{\mega\hertz} (blue) and \SI{1450}{\mega\hertz} (orange)], are very similar to the residuals from averaging over all available frequencies (black, dashed).}
    \label{fig:single_channel_residuals}
\end{figure}
The only remaining possibility is that the `W'-shaped feature is due to a propagation effect, associated with the dips in DM.
A puzzle, though, is that propagation effects tend to be chromatic, with effects typically stronger at lower frequencies, while for \psr, the residuals appears to be roughly achromatic in both amplitude and duration.
Since the change in the profile is seen in all three bands at comparable strength, the observed changes cannot be due to DM smearing, extra scattering tail, and scintillation. 
To confirm the latter, we repeated our analysis on single 1.5-MHz channels, i.e., on a width much smaller than the $\sim\!\SI{23}{\mega\hertz}$ scintillation bandwidth at 1.5 GHz \citep{turnerNANOGrav12Year2021}.
As can be seen in Fig.~\ref{fig:single_channel_residuals}, we recover the profile residuals and find that they are consistent between channels.

This still leaves us with a puzzle.
Fortunately, if the DM change is indeed real, as appears to be the case, there is a clear prediction: in addition to causing dispersive delay, a sufficiently sharp change in DM \textit{must} also cause lensing effects due to geometry.
With baseband data, reprocessed to a sufficiently high-frequency resolution, it should be possible to rule out whether the profile change is intrinsic, or due to additional images, as the latter would cause frequency modulations while the former would not.

\section{Conclusions}
\label{sec:conclusion}
\citet{ATel14642Sustained}'s initial report of the large-amplitude profile change in \psr also reported a dip in DM of $\dmu{-4.3e-3}$, followed by an apparent recovery to a different value, much like the DM time series near the `dips' shown in Fig.~\ref{fig:dms}. The subsequent confirmation of the profile change by \citet{ATel14652Confirmation}, however, attributes the apparent DM dip to a simultaneous frequency-dependent profile change, and instead reports a positive DM jump of $\sim\dmu{1e-3}$ at the same epoch. Indeed, the recent event raises an important question about how to generally disentangle intrinsic frequency-dependent profile change from propagation-induced profile change.

Although the profile change shown in this work is also accompanied by a DM dip, the `W'-shaped residual is only an order 1 per cent change in amplitude relative to the mean profiles away from the events, much smaller than the $\gtrsim 20$ per cent profile change discovered by \citet{ATel14642Sustained}. 
The profile change can, in principle, be responsible for the \dmu{5e-4} dips. However, the needed change has to be highly contrived given the small amplitude, along with the $\lambda^2$ dependence of the ToAs and the achromaticity of the profile residual across the 800-MHz, 1.4-GHz and 2.3-GHz bands.

A sharp enough DM shift will inevitably produce multiple images regardless of the specific lens model, since DM gradient is proportional to the bending angle as per the lens equation (\ref{eq:lenseq}). Therefore, regardless of any intrinsic profile changes, interpreting the frequency-dependent ToAs as due to a DM change gives testable predictions, in contrast to an intrinsic profile change.
\citet{linDiscoveryModellingBroadscale2021} has shown DM variations can successfully predict lensing behavior in the black widow pulsar J2051--0827.
As discussed in Section~\ref{sec:lensing_predictions}, multiple imaging is generically expected from the observed DM at the current cadence and frequencies of observation for \psr. Moreover, the pulsar is expected to dimmer post-dip, but this can be difficult to observe due to scintillation.

These predictions assume fairly smooth DM changes, as is the case with a smooth sheet-like lens analysed in this work.
More complicated lensing behavior can occur if the sheet is not `smooth', in particular when the pulsar is behind the sheet, but the generic predictions from above are still expected.
Since the DM interpretation is testable, one can check for consistency, e.g., by looking for multiple images prior to the dips.
A further prediction assuming a sheet interpretation of DM changes, is that we expect to encounter events that are slow decrease followed by a quick recovery (i.e. sheet folded on the opposite side).

At a frequency resolution of \qty{1.5}{MHz}, the current folded profiles do not have sufficient resolution to find frequency modulations due to multiple imaging.
Finely resolved frequency channels are typically required in the studies of propagation effects.
Recently, \citet{mainMeasuringInterstellarDelays2020a} used rereduced baseband data in order to resolve finely modulating scintillation in PSR~J0613--0200. 
\citet{dolchDeconvolvingPulsarSignals2021} has proposed using cyclic spectroscopy to obtain extremely fine channelization for \textit{NANOGrav}.
To enable more extensive studies of propagation effects, we echo the proposal that future pulsar observations be stored baseband to permit refolding to the necessary resolutions, or in multiple time--frequency resolution trade-offs.

If this profile change is astrophysical in nature, \psr would be one of the few MSPs to show long-term pulse profile change, and the first MSP to show profile change associated with DM variations.

\section*{Acknowledgements}
We are grateful for Michael Lam and the \textit{NANOGrav} collaboration for providing the data used in this analysis.
We thank him, Daniel Baker, the University of Toronto Scintillometry group and the MPIfR pulsar group for useful discussion and comments.
We acknowledge the support of the Natural Sciences and Engineering Research Council of Canada (NSERC, funding reference number RGPIN-2019-067, 523638-201, and 509486). We also receive support from Ontario Research Fund -- Research Excellence Program (ORF-RE), Canadian Institute for Advanced Research (CIFAR), Canadian Foundation for Innovation (CFI), Simons Foundation, Thoth Technology, Inc., and Alexander von Humboldt Foundation. FXL also acknowledges support from the University of Toronto.

\section*{Data availability}
The data underlying this article are publicly available from the \textit{NANOGrav} Collaboration.

%%%%%%%%%%%%%%%%%%%%%%%%%%%%%%%%%%%%%%%%%%%%%%%%%%

%%%%%%%%%%%%%%%%%%%% REFERENCES %%%%%%%%%%%%%%%%%%

\bibliographystyle{mnras}
\bibliography{bibtex} % if your bibtex file is called example.bib

\begin{thebibliography}{}
\makeatletter
\relax
\def\mn@urlcharsother{\let\do\@makeother \do\$\do\&\do\#\do\^\do\_\do\%\do\~}
\def\mn@doi{\begingroup\mn@urlcharsother \@ifnextchar [ {\mn@doi@}
  {\mn@doi@[]}}
\def\mn@doi@[#1]#2{\def\@tempa{#1}\ifx\@tempa\@empty \href
  {http://dx.doi.org/#2} {doi:#2}\else \href {http://dx.doi.org/#2} {#1}\fi
  \endgroup}
\def\mn@eprint#1#2{\mn@eprint@#1:#2::\@nil}
\def\mn@eprint@arXiv#1{\href {http://arxiv.org/abs/#1} {{\tt arXiv:#1}}}
\def\mn@eprint@dblp#1{\href {http://dblp.uni-trier.de/rec/bibtex/#1.xml}
  {dblp:#1}}
\def\mn@eprint@#1:#2:#3:#4\@nil{\def\@tempa {#1}\def\@tempb {#2}\def\@tempc
  {#3}\ifx \@tempc \@empty \let \@tempc \@tempb \let \@tempb \@tempa \fi \ifx
  \@tempb \@empty \def\@tempb {arXiv}\fi \@ifundefined
  {mn@eprint@\@tempb}{\@tempb:\@tempc}{\expandafter \expandafter \csname
  mn@eprint@\@tempb\endcsname \expandafter{\@tempc}}}

\bibitem[\protect\citeauthoryear{Alam et~al.,}{Alam
  et~al.}{2020}]{alamNANOGrav12Yr2020}
Alam M.~F.,  et~al., 2020, \mn@doi [The Astrophysical Journal Supplement
  Series] {10.3847/1538-4365/abc6a0}, 252, 4

\bibitem[\protect\citeauthoryear{Antoniadis et~al.,}{Antoniadis
  et~al.}{2013}]{antoniadisMassivePulsarCompact2013}
Antoniadis J.,  et~al., 2013, \mn@doi [Science] {10.1126/science.1233232}, 340,
  448

\bibitem[\protect\citeauthoryear{Archibald et~al.,}{Archibald
  et~al.}{2018}]{archibaldUniversalityFreeFall2018}
Archibald A.~M.,  et~al., 2018, \mn@doi [Nature] {10.1038/s41586-018-0265-1},
  559, 73

\bibitem[\protect\citeauthoryear{Arzoumanian et~al.,}{Arzoumanian
  et~al.}{2018}]{arzoumanianNANOGrav11yearData2018}
Arzoumanian Z.,  et~al., 2018, \mn@doi [The Astrophysical Journal Supplement
  Series] {10/ggpw4x}, 235, 37

\bibitem[\protect\citeauthoryear{Arzoumanian et~al.,}{Arzoumanian
  et~al.}{2020}]{arzoumanianNANOGrav12Yr2020}
Arzoumanian Z.,  et~al., 2020, \mn@doi [The Astrophysical Journal]
  {10.3847/2041-8213/abd401}, 905, L34

\bibitem[\protect\citeauthoryear{Backer, Wong  \& Valanju}{Backer
  et~al.}{2000}]{backerPlasmaPrismModel2000}
Backer D.~C.,  Wong T.,   Valanju J.,  2000, \mn@doi [The Astrophysical
  Journal] {10/bhrqzz}, 543, 740

\bibitem[\protect\citeauthoryear{Bassa et~al.,}{Bassa
  et~al.}{2016}]{bassaLEAPLargeEuropean2016}
Bassa C.~G.,  et~al., 2016, \mn@doi [Monthly Notices of the Royal Astronomical
  Society] {10/ggrkrx}, 456, 2196

\bibitem[\protect\citeauthoryear{Brisken, Macquart, Gao, Rickett, Coles,
  Deller, Tingay  \& West}{Brisken et~al.}{2010}]{brisken100MasResolution2010}
Brisken W.~F.,  Macquart J.-P.,  Gao J.~J.,  Rickett B.~J.,  Coles W.~A.,
  Deller A.~T.,  Tingay S.~J.,   West C.~J.,  2010, \mn@doi [The Astrophysical
  Journal] {10/bmkhmb}, 708, 232

\bibitem[\protect\citeauthoryear{Brook et~al.,}{Brook
  et~al.}{2018}]{brookNANOGrav11yearData2018}
Brook P.~R.,  et~al., 2018, \mn@doi [The Astrophysical Journal] {10/gft2mh},
  868, 122

\bibitem[\protect\citeauthoryear{Cognard \& Backer}{Cognard \&
  Backer}{2004}]{cognardMicroglitchMillisecondPulsar2004}
Cognard I.,  Backer D.~C.,  2004, \mn@doi [The Astrophysical Journal Letters]
  {10.1086/424692}, 612, L125

\bibitem[\protect\citeauthoryear{Coles, Rickett, Gao, Hobbs  \& Verbiest}{Coles
  et~al.}{2010}]{colesScatteringPulsarRadio2010}
Coles W.~A.,  Rickett B.~J.,  Gao J.~J.,  Hobbs G.,   Verbiest J. P.~W.,  2010,
  \mn@doi [The Astrophysical Journal] {10.1088/0004-637X/717/2/1206}, 717, 1206

\bibitem[\protect\citeauthoryear{Coles et~al.,}{Coles
  et~al.}{2015}]{colesPulsarObservationsExtreme2015}
Coles W.~A.,  et~al., 2015, \mn@doi [The Astrophysical Journal]
  {10.1088/0004-637X/808/2/113}, 808, 113

\bibitem[\protect\citeauthoryear{Cordes \& Wolszczan}{Cordes \&
  Wolszczan}{1986}]{cordesMultipleImagingPulsars1986}
Cordes J.~M.,  Wolszczan A.,  1986, \mn@doi [The Astrophysical Journal]
  {10.1086/184722}, 307, L27

\bibitem[\protect\citeauthoryear{Cordes, Wasserman, Hessels, Lazio, Chatterjee
  \& Wharton}{Cordes et~al.}{2017}]{cordesLensingFastRadio2017}
Cordes J.~M.,  Wasserman I.,  Hessels J. W.~T.,  Lazio T. J.~W.,  Chatterjee
  S.,   Wharton R.~S.,  2017, \mn@doi [The Astrophysical Journal] {10/gf6nsn},
  842, 35

\bibitem[\protect\citeauthoryear{Desvignes et~al.,}{Desvignes
  et~al.}{2016}]{desvignesHighprecisionTiming422016}
Desvignes G.,  et~al., 2016, \mn@doi [Monthly Notices of the Royal Astronomical
  Society] {10.1093/mnras/stw483}, 458, 3341

\bibitem[\protect\citeauthoryear{Dolch et~al.,}{Dolch
  et~al.}{2021}]{dolchDeconvolvingPulsarSignals2021}
Dolch T.,  et~al., 2021, \mn@doi [The Astrophysical Journal]
  {10.3847/1538-4357/abf48b}, 913, 98

\bibitem[\protect\citeauthoryear{Espinoza, Lyne, Stappers  \& Kramer}{Espinoza
  et~al.}{2011}]{espinozaStudy315Glitches2011}
Espinoza C.~M.,  Lyne A.~G.,  Stappers B.~W.,   Kramer M.,  2011, \mn@doi
  [Monthly Notices of the Royal Astronomical Society] {10/c6rnc4}, 414, 1679

\bibitem[\protect\citeauthoryear{Goncharov et~al.,}{Goncharov
  et~al.}{2021}]{goncharovIdentifyingMitigatingNoise2021}
Goncharov B.,  et~al., 2021, \mn@doi [Monthly Notices of the Royal Astronomical
  Society] {10.1093/mnras/staa3411}, 502, 478

\bibitem[\protect\citeauthoryear{Jones et~al.,}{Jones
  et~al.}{2017}]{jonesNANOGravNineyearData2017}
Jones M.~L.,  et~al., 2017, \mn@doi [The Astrophysical Journal] {10/gd4vr8},
  841, 125

\bibitem[\protect\citeauthoryear{Kerr et~al.,}{Kerr
  et~al.}{2020}]{kerrParkesPulsarTiming2020}
Kerr M.,  et~al., 2020, \mn@doi [Publications of the Astronomical Society of
  Australia] {10.1017/pasa.2020.11}, 37, e020

\bibitem[\protect\citeauthoryear{Lam et~al.,}{Lam
  et~al.}{2018}]{lamSecondChromaticTiming2018}
Lam M.~T.,  et~al., 2018, \mn@doi [The Astrophysical Journal] {10/gdtgwm}, 861,
  132

\bibitem[\protect\citeauthoryear{Lentati et~al.,}{Lentati
  et~al.}{2017}]{lentatiWidebandProfileDomain2017}
Lentati L.,  et~al., 2017, \mn@doi [Monthly Notices of the Royal Astronomical
  Society] {10/f9453q}, 466, 3706

\bibitem[\protect\citeauthoryear{Levkov, Panin  \& Tkachev}{Levkov
  et~al.}{2020}]{levkovPeriodicStructureFRB2020}
Levkov D.~G.,  Panin A.~G.,   Tkachev I.~I.,  2020, arXiv:2010.15145 [astro-ph,
  physics:hep-ph]

\bibitem[\protect\citeauthoryear{Liddle}{Liddle}{2007}]{liddleInformationCriteriaAstrophysical2007}
Liddle A.~R.,  2007, \mn@doi [Monthly Notices of the Royal Astronomical
  Society] {10.1111/j.1745-3933.2007.00306.x}, 377, L74

\bibitem[\protect\citeauthoryear{Lin, Masui, Pen  \& Peterson}{Lin
  et~al.}{2018}]{linImprovedPulsarTiming2018}
Lin H.-H.,  Masui K.,  Pen U.-L.,   Peterson J.~B.,  2018, \mn@doi [Monthly
  Notices of the Royal Astronomical Society] {10/gdbzvx}, 475, 1323

\bibitem[\protect\citeauthoryear{Lin, Main, Verbiest, Kramer  \&
  Shaifullah}{Lin et~al.}{2021}]{linDiscoveryModellingBroadscale2021}
Lin F.~X.,  Main R.~A.,  Verbiest J. P.~W.,  Kramer M.,   Shaifullah G.,  2021,
  \mn@doi [Monthly Notices of the Royal Astronomical Society]
  {10.1093/mnras/stab1811}, 506, 2824

\bibitem[\protect\citeauthoryear{Liu et~al.,}{Liu
  et~al.}{2016}]{liuVariabilityPolarimetryTiming2016}
Liu K.,  et~al., 2016, \mn@doi [Monthly Notices of the Royal Astronomical
  Society] {10/f9rxmj}, 463, 3239

\bibitem[\protect\citeauthoryear{Lyne, Hobbs, Kramer, Stairs  \& Stappers}{Lyne
  et~al.}{2010}]{lyneSwitchedMagnetosphericRegulation2010}
Lyne A.,  Hobbs G.,  Kramer M.,  Stairs I.,   Stappers B.,  2010, \mn@doi
  [Science] {10/dzktfw}, 329, 408

\bibitem[\protect\citeauthoryear{Mahajan, {van Kerkwijk}, Main  \& Pen}{Mahajan
  et~al.}{2018}]{mahajanModeChangingGiant2018}
Mahajan N.,  {van Kerkwijk} M.~H.,  Main R.,   Pen U.-L.,  2018, \mn@doi [The
  Astrophysical Journal Letters] {10.3847/2041-8213/aae713}, 867, L2

\bibitem[\protect\citeauthoryear{Main et~al.,}{Main
  et~al.}{2020}]{mainMeasuringInterstellarDelays2020a}
Main R.~A.,  et~al., 2020, \mn@doi [Monthly Notices of the Royal Astronomical
  Society] {10.1093/mnras/staa2955}, 499, 1468

\bibitem[\protect\citeauthoryear{Marthi et~al.,}{Marthi
  et~al.}{2021}]{marthiScintillationPSRB15082021a}
Marthi V.~R.,  et~al., 2021, \mn@doi [Monthly Notices of the Royal Astronomical
  Society] {10.1093/mnras/stab1970}, 506, 5160

\bibitem[\protect\citeauthoryear{McKee et~al.,}{McKee
  et~al.}{2016}]{mckeeGlitchMillisecondPulsar2016}
McKee J.~W.,  et~al., 2016, \mn@doi [Monthly Notices of the Royal Astronomical
  Society] {10.1093/mnras/stw1442}, 461, 2809

\bibitem[\protect\citeauthoryear{Meyers \& {The CHIME/Pulsar
  Collaboration}}{Meyers \& {The CHIME/Pulsar
  Collaboration}}{2021}]{ATel14652Confirmation}
Meyers B.,  {The CHIME/Pulsar Collaboration} 2021, {{ATel}} \#14652:
  {{Confirmation}} of a Change in the Emission Properties of {{PSR
  J1713}}+0747, https://www.astronomerstelegram.org/?read=14652

\bibitem[\protect\citeauthoryear{Os{\l}owski, {van Straten}, Hobbs, Bailes  \&
  Demorest}{Os{\l}owski et~al.}{2011}]{oslowskiHighSignaltonoiseRatio2011}
Os{\l}owski S.,  {van Straten} W.,  Hobbs G.~B.,  Bailes M.,   Demorest P.,
  2011, \mn@doi [Monthly Notices of the Royal Astronomical Society]
  {10/bmghmr}, 418, 1258

\bibitem[\protect\citeauthoryear{Padmanabh, Barr, Champion, Karuppusamy,
  Kramer, Jessner  \& Lazarus}{Padmanabh
  et~al.}{2021}]{padmanabhRevisitingProfileInstability2021}
Padmanabh P.~V.,  Barr E.~D.,  Champion D.~J.,  Karuppusamy R.,  Kramer M.,
  Jessner A.,   Lazarus P.,  2021, \mn@doi [Monthly Notices of the Royal
  Astronomical Society] {10.1093/mnras/staa3174}, 500, 1178

\bibitem[\protect\citeauthoryear{Pen \& Levin}{Pen \&
  Levin}{2014}]{penPulsarScintillationsCorrugated2014}
Pen U.-L.,  Levin Y.,  2014, \mn@doi [Monthly Notices of the Royal Astronomical
  Society] {10/f6c484}, 442, 3338

\bibitem[\protect\citeauthoryear{Phillips \& Wolszczan}{Phillips \&
  Wolszczan}{1991}]{phillipsTimeVariabilityPulsar1991}
Phillips J.~A.,  Wolszczan A.,  1991, \mn@doi [The Astrophysical Journal]
  {10/b3c9qj}, 382, L27

\bibitem[\protect\citeauthoryear{Putney \& Stinebring}{Putney \&
  Stinebring}{2006}]{putneyMultipleScintillationArcs2006}
Putney M.~L.,  Stinebring D.~R.,  2006, Chinese Journal of Astronomy and
  Astrophysics Supplement, 6, 233

\bibitem[\protect\citeauthoryear{Ransom et~al.,}{Ransom
  et~al.}{2019}]{ransomNANOGravProgramGravitational2019b}
Ransom S.,  et~al., 2019, Bulletin of the American Astronomical Society, 51,
  195

\bibitem[\protect\citeauthoryear{Shannon \& Cordes}{Shannon \&
  Cordes}{2012}]{shannonPulseIntensityModulation2012}
Shannon R.~M.,  Cordes J.~M.,  2012, \mn@doi [The Astrophysical Journal]
  {10/ggvb2h}, 761, 64

\bibitem[\protect\citeauthoryear{Shannon et~al.,}{Shannon
  et~al.}{2016}]{shannonDisturbanceMillisecondPulsar2016}
Shannon R.~M.,  et~al., 2016, \mn@doi [The Astrophysical Journal] {10/ggc8dw},
  828, L1

\bibitem[\protect\citeauthoryear{Simard \& Pen}{Simard \&
  Pen}{2018}]{simardPredictingPulsarScintillation2018}
Simard D.,  Pen U.-L.,  2018, \mn@doi [Monthly Notices of the Royal
  Astronomical Society] {10/gd2s6r}, 478, 983

\bibitem[\protect\citeauthoryear{Stinebring}{Stinebring}{2013}]{stinebringEffectsInterstellarMedium2013}
Stinebring D.,  2013, \mn@doi [Classical and Quantum Gravity]
  {10.1088/0264-9381/30/22/224006}, 30, 224006

\bibitem[\protect\citeauthoryear{Susobhanan et~al.,}{Susobhanan
  et~al.}{2021}]{susobhananPintaUGMRTData2021}
Susobhanan A.,  et~al., 2021, \mn@doi [Publications of the Astronomical Society
  of Australia] {10.1017/pasa.2021.12}, 38, e017

\bibitem[\protect\citeauthoryear{Taylor}{Taylor}{1992}]{taylorPulsarTimingRelativistic1992}
Taylor J.~H.,  1992, \mn@doi [Philosophical Transactions of the Royal Society
  of London Series A] {10/fg4bh7}, 341, 117

\bibitem[\protect\citeauthoryear{{Thulasiram} \& {Lin}}{{Thulasiram} \&
  {Lin}}{2021}]{2021arXiv210513316T}
{Thulasiram} P.,  {Lin} H.-H.,  2021, arXiv e-prints, \href
  {https://ui.adsabs.harvard.edu/abs/2021arXiv210513316T} {p. arXiv:2105.13316}

\bibitem[\protect\citeauthoryear{Turner et~al.,}{Turner
  et~al.}{2021}]{turnerNANOGrav12Year2021}
Turner J.~E.,  et~al., 2021, \mn@doi [The Astrophysical Journal]
  {10.3847/1538-4357/abfafe}, 917, 10

\bibitem[\protect\citeauthoryear{Verbiest, Oslowski  \&
  {Burke-Spolaor}}{Verbiest et~al.}{2021}]{verbiestPulsarTimingArray2021}
Verbiest J. P.~W.,  Oslowski S.,   {Burke-Spolaor} S.,  2021, arXiv:2101.10081
  [astro-ph]

\bibitem[\protect\citeauthoryear{Wucknitz}{Wucknitz}{2019}]{wucknitzImagingPulsarEchoes2019}
Wucknitz O.,  2019, arXiv e-prints, 1904, arXiv:1904.11347

\bibitem[\protect\citeauthoryear{Xu et~al.,}{Xu
  et~al.}{2021}]{ATel14642Sustained}
Xu H.,  et~al., 2021, {{ATel}} \#14642: {{A}} Sustained Pulse Shape Change in
  {{PSR J1713}}+0747 Possibly Associated with Timing and {{DM}} Events,
  https://www.astronomerstelegram.org/?read=14642

\bibitem[\protect\citeauthoryear{Zhu et~al.,}{Zhu
  et~al.}{2015}]{zhuTestingTheoriesGravitation2015}
Zhu W.~W.,  et~al., 2015, \mn@doi [The Astrophysical Journal] {10/ggth73}, 809,
  41

\bibitem[\protect\citeauthoryear{Zhu et~al.,}{Zhu
  et~al.}{2019}]{zhuTestsGravitationalSymmetries2019}
Zhu W.~W.,  et~al., 2019, \mn@doi [Monthly Notices of the Royal Astronomical
  Society] {10/ggth72}, 482, 3249

\makeatother
\end{thebibliography}
%%%%%%%%%%%%%%%%%%%%%%%%%%%%%%%%%%%%%%%%%%%%%%%%%%

%%%%%%%%%%%%%%%%% APPENDICES %%%%%%%%%%%%%%%%%%%%%

\appendix
\section{Supplementary figures}
\begin{figure*}
  \begin{center}
     \resizebox{\textwidth}{!}{\input{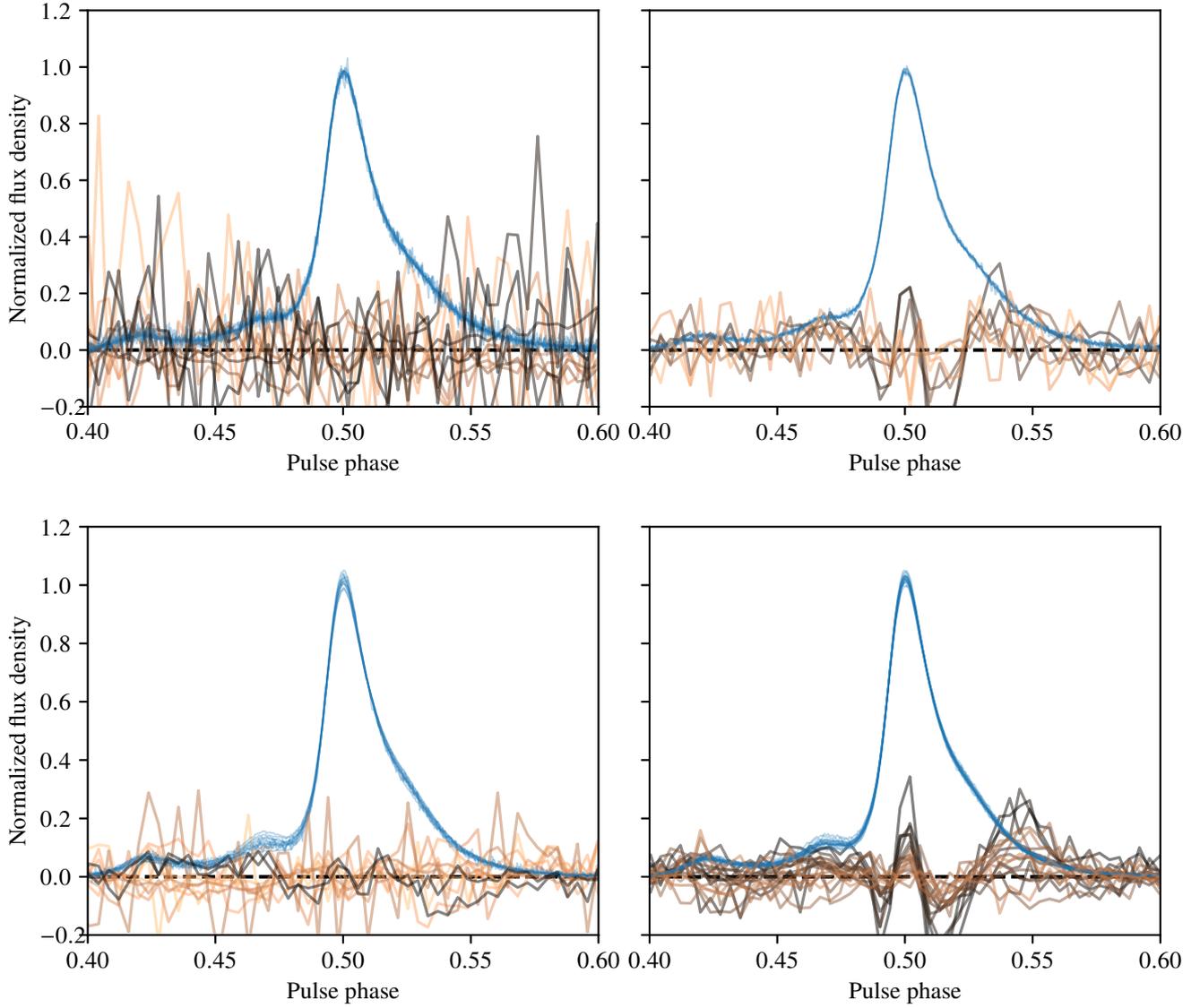}}
  \end{center}
  \caption{
    Profile residuals compared with the integrated profile.
    The integrated profiles of each observation are shown in blue and the profile residuals are in brown, with darker brown used for times closer to the DM event.
    For clarity, only relatively high S/N ratio profiles are shown, and profile residuals are rebinned by a factor 8 in pulse phase, and scaled by factor 30.
    \textit{Top}: \textit{Green Bank} 1.5-GHz observations around the first event, with 11 profiles and residuals from the 250 d prior to MJD 54755 on the left, and 11 profiles and residuals in the 250 d after MJD 54755 on the right.
    \textit{Bottom}: \textit{Arecibo} 2.3-GHz observations around the second event, with 11 profiles and residuals in the 150 d prior to MJD 57512 on the left, and 18 profiles and residuals in the 150 d after MJD 57512 on the right.
  }
  \label{fig:profilesandresiduals_simple}
\end{figure*}

We provide a few more figures to help understand the profile changes.
First, Fig.~\ref{fig:profilesandresiduals_simple} shows a different view of the profiles, both away and at the events, to show that even though the residuals associated with the DM events are small, they are significant.

\begin{figure*}
  \begin{center}
     \resizebox{\textwidth}{!}{\input{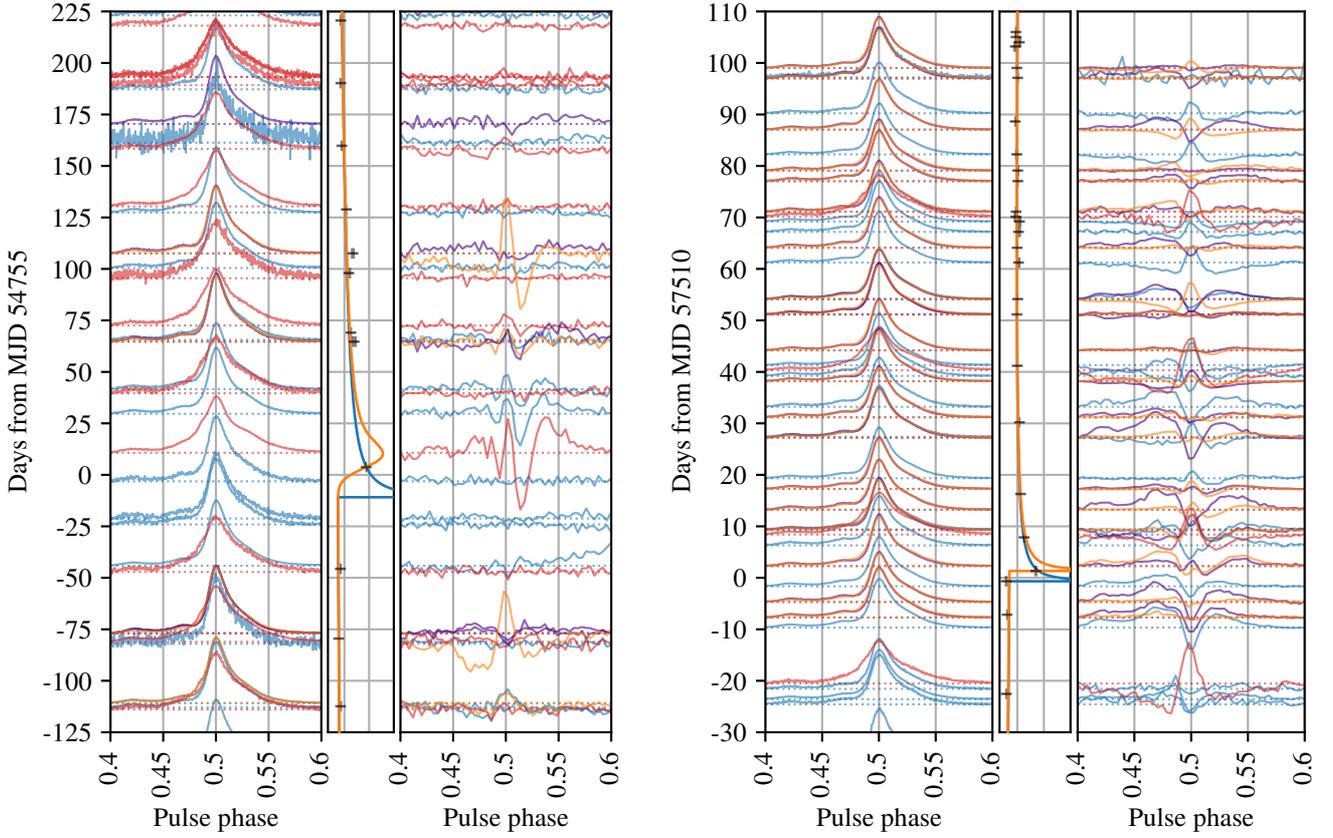}}
  \end{center}
  \caption{
    Like Fig.~\ref{fig:profilesandresiduals}, but showing profile residuals relative to the mean profile, and with the residuals in the rightmost panel scaled by only 15-times since otherwise they would overlap too much.
    For the first event, one can still make out the characteristic W-shaped features in the residuals for a few observations after the DM dip, but for the second event, the broadening mode dominates the profile residuals.
   }
   \label{fig:profilesandresiduals_meansubtraction}
\end{figure*}

Second, Fig.~\ref{fig:profilesandresiduals_meansubtraction} shows the effect of only subtracting a static mean profile, instead both the mean and the broadening mode.
For the first event, the residual `W'-shaped feature is still visible, but in the second event it is hidden by the broadening mode.

\begin{figure*}
  \begin{center}
     \resizebox{\textwidth}{!}{\input{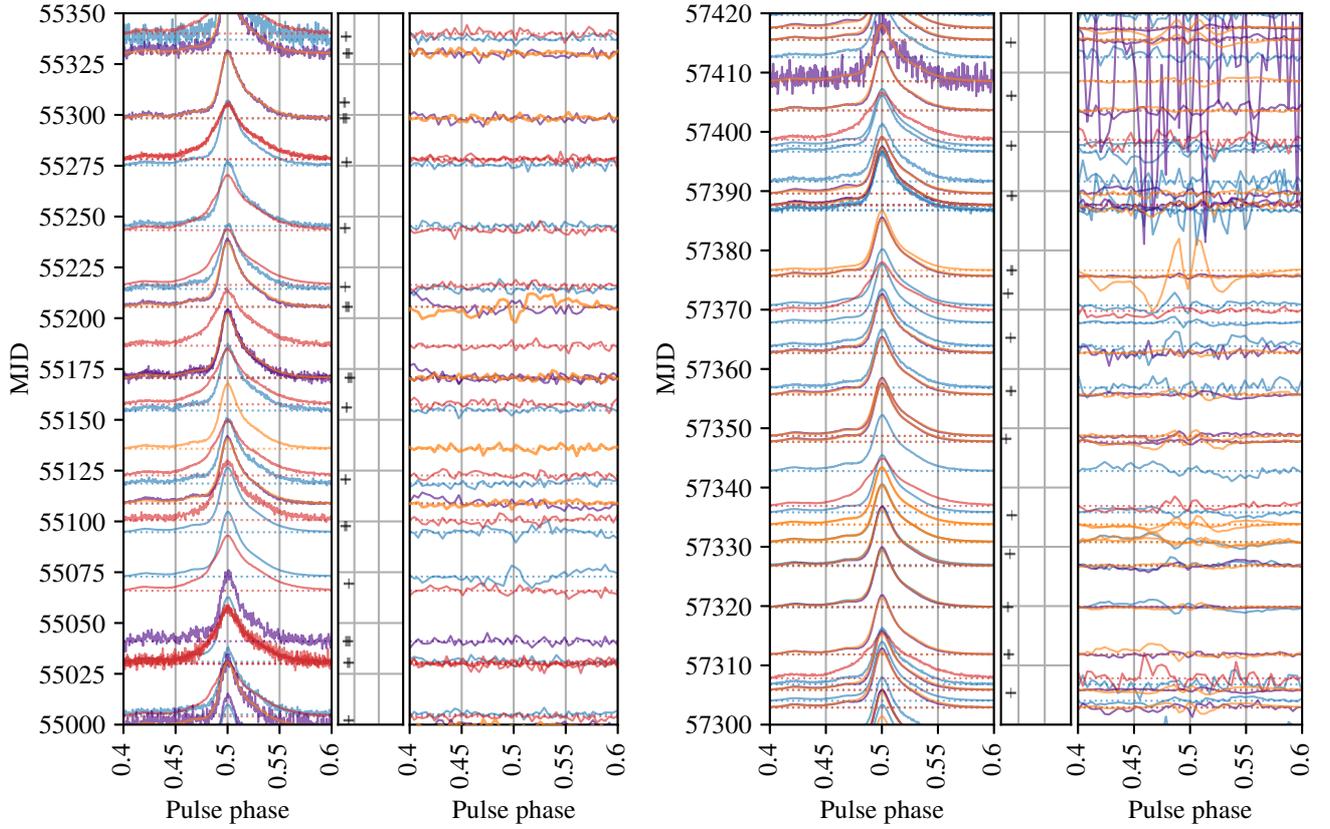}}
  \end{center}
  \caption{
    Like Fig.~\ref{fig:profilesandresiduals}, but for times outside the event times but with the same backends (left: ASP/GASP, right: PUPPI/GUPPI).
    The residuals are for profiles with the first two \textit{V}-modes removed, and show no systemic evolution, unlike for those associated with the DM events.
    Note the effect of undersubtracting the broadening mode at \textit{Arecibo} L-band (orange) at, e.g. $\sim$MJD 57375.
   }
   \label{fig:profilesandresiduals_outside}
\end{figure*}

Third, Fig.~\ref{fig:profilesandresiduals_outside} again shows residuals with both the mean profile and the broadening mode subtracted, but now at times far away from the DM events.
At those times, no systematic evolution of profile residual can be seen.

\bsp	% typesetting comment
\label{lastpage}
\end{document}